%% file: final_draft_v2.tex
\documentclass[a4paper,11pt]{article}

\usepackage{jheppub} 
\usepackage{graphicx,color}
\usepackage{amsmath}
\usepackage{slashed}
\usepackage{multirow}
\usepackage{autobreak}
\usepackage[normalem]{ulem}
\usepackage{hyperref}
\usepackage{cleveref}
\usepackage{slashed}
\usepackage{amssymb}
\usepackage{amsmath}
\usepackage{mathrsfs}
\usepackage{pifont}
\usepackage{epsfig}
\usepackage{array}
\usepackage{mathtools}
\usepackage{caption}
\usepackage[utf8]{inputenc}
\usepackage{mathtools, nccmath}
\usepackage{float}
\usepackage[section]{placeins}
\usepackage[utf8]{inputenc}
\usepackage[T1]{fontenc}
\usepackage{lmodern}
\usepackage[x11names]{xcolor}
\usepackage{framed}
\colorlet{shadecolor}{blue!10}

\SetSymbolFont{largesymbols}{bold}{OMX}{txex}{b}{n}

\DeclareMathAlphabet{\mathpzc}{OT1}{pzc}{m}{it}

\newcommand{\be}{\begin{eqnarray*}}
	\newcommand{\ee}{\end{eqnarray*}}

\usepackage{parskip}

\newcommand{\ba}{\begin{array}}
	\newcommand{\ea}{\end{array}}
\newcommand{\bd}{\begin{displaymath}}
	\newcommand{\ed}{\end{displaymath}}
\newcommand{\besub}{\begin{subequations}}
	\newcommand{\eesub}{\end{subequations}}

\def\q2 {q^2}

\def\bt{\begin{table}}
	\def\et{\end{table}}

\input mylatex

	\title{\boldmath Optimal Sensitivity of Anomalous Charged Triple Gauge Couplings through $W$ boson helicity at the $e^+e^-$ colliders}

\author[a,b]{Sahabub Jahedi,}
\author[c]{Jayita Lahiri}
\author[d]{and Amir Subba}

\affiliation[a]{State Key Laboratory of Nuclear Physics and Technology, Institute of Quantum Matter, South China Normal 
	University, Guangzhou 510006, China}
\affiliation[b]{Guangdong Basic Research Center of Excellence for Structure and Fundamental Interactions of Matter, Guangdong 
	Provincial Key Laboratory of Nuclear Science, Guangzhou 510006, China}
\affiliation[c]{Institut f{\"u}r Theoretische Physik, Universit{\"a}t Hamburg, Luruper Chaussee 149, 22761 Hamburg, Germany}
\affiliation[d]{Department of Physical Sciences, Indian Institute of Science Education and Research Kolkata, Mohanpur 74126, India}

\emailAdd{sahabub@m.scnu.edu.cn}
\emailAdd{jayita.lahiri@desy.de}
\emailAdd{as19rs008@iiserkol.ac.in}

\abstract{ We study the estimation of anomalous charged triple gauge couplings (cTGCs) parameterized in a model-independent Standard Model effective field theory (SMEFT) framework via $WW$ production followed by semi-leptonic decay at the $e^+e^-$ colliders. The anomalous part of cTGCs, $WWV~(V=\gamma,Z)$, are given in terms of Wilson coefficients of three CP-conserving and two CP-violating dimension-6 operators in the HISZ basis. We adopt the optimal observable technique (OOT) to extract the sensitivity of these anomalous couplings and compare it with the latest experimental limits on anomalous couplings studied at the LHC. The limits on the anomalous couplings obtained via OOT are significantly tighter than the ones obtained using standard $\chi^2$ analysis. The impact of different combinations of the helicity of $W$ boson pair in determining the optimal sensitivity is analyzed. The constraints on CP-violating operators from the electron electric dipole moment (EDM) are also discussed.}


\begin{document}

\maketitle
\flushbottom

\section{Introduction}
\label{sec:intro}
\noindent
The discovery of the Higgs boson at the large hadron collider (LHC) \cite{Aad:2012tfa,Chatrchyan:2012ufa} has marked a major triumph for the Standard Model (SM) as the most credible theory to describe nature. However, this milestone also brought the pressing question to the forefront: is there any physics beyond the Standard Model (BSM). In the absence of any decisive signal of new physics (NP), it is essential in the upcoming years to pin down the electroweak symmetry breaking (EWSB) scenario by measuring the SM couplings with unprecedented precision. The gauge interactions within the electroweak sector are an integral part of this quest. The non-abelian $SU(2)_L \times U(1)_Y$ group suggests that self-interaction among gauge bosons exists within the EW sector. Therefore, the measurement of self-interactions of gauge bosons such as charged triple gauge couplings~(cTGCs) $WWV~(V=\gamma, Z$) and/or quartic gauge couplings~(QGCs) have to be very precise, which is yet to be done. Any deviation in the measurement of gauge boson self-interactions from the SM predictions would be a clear indication of NP. Exploring NP search via $WWV$ couplings may also provide an additional probe of CP violation, which can explain electroweak baryogenesis. The precise measurement of gauge boson self-interactions also becomes important to test whether the gauge symmetry is realized linearly or non-linearly in the low energy effective field theory.

The precision measurement of anomalous cTGCs points us toward a lepton collider. This choice is driven by two key advantages: first, the elimination of QCD backgrounds and the reduction of the parton distribution function (PDF) uncertainties, both of which are inherent challenges at hadron colliders; second, the ability to utilize initial beam polarization, which can enhance the NP signal and/or suppress the SM backgrounds. Therefore, we consider a proposed electron-positron collider, {\it e.g.,} compact linear collider (CLIC) \cite{CLICdp:2018cto} to determine the sensitivity of cTGCs at the baseline design of the CLIC and compare existing bounds given by the CMS experiments \cite{CMS:2019ppl,CMS:2021icx,CMS:2021foa} at the LHC. 

Di-boson production at colliders has been receiving a lot of attention since the past as this is sensitive to modification of triple gauge couplings~(TGC). In this regard, $W$-boson pair production plays a crucial role to estimate the sensitivity of anomalous cTGCs. The phenomenology of cTGCs through $W$ boson pair production has been studied at the hadron colliders \cite{Baur:1987mt,Bian:2015zha,Butter:2016cvz,Azatov:2017kzw,Baglio:2017bfe,Li:2017esm,Chiesa:2018lcs,Baglio:2018bkm,Bhatia:2018ndx,Rahaman:2019lab,Azatov:2019xxn,Tizchang:2020tqs}, lepton colliders \cite{Gaemers:1978hg,Bilchak:1984ur,Hagiwara:1986vm,Chang:1993vv,Hagiwara:1992eh,Choudhury:1996ni,Choudhury:1999fz,Buchalla:2013wpa,Berthier:2016tkq,Zhang:2016zsp,Rahaman:2019mnz,Subba:2022czw,Subba:2023rpm,Subba:2024ojy}, and electron-ion colliders \cite{Cakir:2014swa,Li:2017kfk,Gutierrez-Rodriguez:2019hek,Koksal:2019oqt,Spor:2020rig,Spor:2021dhf}. Pair production of $W$-boson at the $e^+e^-$ colliders is governed by neutrino-mediated $t$-channel diagram and $\gamma$ and $Z$ mediated $s$-channel diagram. In this work, we will study the estimation of anomalous cTGCs through $WW$ production at the CLIC with center-of-mass (CM) energy ($\sqrt{s}$) = 3 TeV and integrated luminosity ($\mathfrak{L}_{\text{int}}$) = 1000 $\rm{fb}^{-1}$, using different choice of beam polarization for different $WW$ helicity configurations. The analysis follows the mathematical framework described in \cite{Degrande:2012wf}, which we describe in the next section. The modification of cTGC vertices appears via $s$-channel mediation. The estimation of future sensitivity of cTGCs is done using the optimal observable technique (OOT) \cite{Atwood:1991ka,Davier:1992nw,Diehl:1993br,Gunion:1996vv}, a statistical method to determine the statistical sensitivity in an economical way. The OOT has been used to perform statistical analysis in the context of Higgs couplings \cite{Hagiwara:2000tk,Dutta:2008bh}, top-quark couplings \cite{Grzadkowski:1996pc,Grzadkowski:1997cj,Grzadkowski:1998bh,Grzadkowski:1999kx,Grzadkowski:2000nx,Bhattacharya:2023mjr}, neutral triple gauge couplings \cite{Jahedi:2022duc,Jahedi:2023myu}, and $Z$ couplings with vector-like leptons \cite{Bhattacharya:2021ltd,Bhattacharya:2023zln,Jahedi:2024koa} at the $e^+e^-$ colliders. Using the OOT, search for NP through $t\bar{t}$ production has been performed at the $\gamma \gamma$ colliders \cite{Grzadkowski:2003tf,Grzadkowski:2004iw,Grzadkowski:2005ye}. This statistical method is also applied to explore the CP properties of the Higgs boson at the LHC \cite{Gunion:1998hm}, the $e \gamma$ collider \cite{Cao:2006pu}, the muon collider \cite{Hioki:2007jc}, and in the field of flavor physics \cite{Bhattacharya:2015ida,Calcuttawala:2017usw,Calcuttawala:2018wgo,Bhattacharya:2023beo,Jahedi:2024kvi}. Determination of future sensitivity of anomalous cTGCs using different helicity combination of $W$ boson pair has not been done yet, to the best of our knowledge. We consider different helicity combinations of $W$ boson pair to determine the sensitivity of anomalous cTGCs using the OOT and compare them with standard $\chi^2$ analysis at the CLIC as well as existing bounds from the LHC. Furthermore, we compare our results with the latest bounds from the electron electric dipole moment (EDM) experiment \cite{ACME:2018yjb} in the case of the CP-violating effective couplings.

This paper is organized as follows: in Sec.~\ref{sec:eft}, we discuss the anomalous parameters within the SMEFT framework. The SM and SMEFT contributions to the $WW$ production are discussed in Sec.~\ref{sec:ww.prod}.  We then give a brief description of the OOT in Sec.~\ref{sec:oot} and discuss the limits obtained on five independent dimension-6 effective couplings via OOT for different helicity combinations of two $W$ bosons. In Sec.~\ref{sec:edm}, we discuss the limits on the CP-odd couplings through the EDM experiments, and finally, we conclude in Sec.~\ref{sec:con}.
%
\section{Anomalous charged triple gauge couplings}
\label{sec:eft}
The most general couplings of two charged vector bosons with a neutral vector boson, $WW\gamma/Z$, considering $U(1)_{\mathrm{EM}}$ gauge and Lorentz invariance, can be derived from the following effective Lagrangian~\cite{Hagiwara:1986vm}
\begin{align}
	\label{eq:hag}
	\nonumber
	\mathcal{L}_{WWV}&=i g_{WWV} (g^V_1(W_{\mu \nu}^+ W^{-\mu}-W^{+\mu}W^{-}_{\mu \nu})V^{\nu}+i g_4^V W^{+}_{\mu} W^{-}_{\nu}(\partial^{\mu} V^{\nu}+\partial^{\nu} V^{\mu})\\\nonumber
	&-ig^V_5\epsilon^{\mu \nu \rho \sigma}(
	W_{\mu}^+ \partial_{\rho} W_{\nu}^- - \partial_{\rho} W_{\mu}^+ W_{\nu}^-)V_{\sigma}+\frac{\lambda^V}{m_W^2}W_{\mu}^{+\nu} W_{\nu}^{-\rho} V_{\rho}^{\mu}+\frac{\widetilde{\lambda}^{V}}{m_W^2}W_{\mu}^{+\nu} W_{\nu}^{-\rho} \widetilde{V}_{\rho}^{\mu}\\
	&+\kappa^V W^{+}_{\mu} W_{\nu}^{-} V^{\mu \nu}
	+\widetilde{\kappa}^{V} W^{+}_{\mu} W_{\nu}^{-} \widetilde{V}^{\mu \nu}),
\end{align} 
where $W^\pm_{\mu\nu} = \partial_\mu W^\pm_\nu - \partial_\nu W^\pm_\mu$ and the dual field is defined as $\widetilde{V}_{\mu\nu}=1/2\epsilon_{\mu\nu\rho\sigma}V^{\rho\sigma}$, with the Levi-Civita tensor following the standard convention, $i.e.$, $\epsilon_{0123}=1$. The $SU(2)_L$ coupling constants related to photon and $Z$ boson are, $g_{WW \gamma}=-g \sin\theta_W$ and $g_{WWZ}=-g \cos\theta_W$, where $\theta_W$ is the weak mixing angle. The couplings $g^V_1$, $\lambda^V$ and $\kappa^V$ are CP-even, while $\tilde{\lambda}^{V}$ and $\tilde{\kappa}^{V}$ are CP-odd. Within the SM, $g_1^\gamma = g_1^Z=1$ and all other couplings are explicitly zero. Electromagnetic gauge invariance implies that any NP contribution to $g_1^\gamma$ should vanish. Furthermore, $g_4^{\gamma}$ and $g_5^{\gamma}$ should also vanish. In the momentum space, $W(q_1)W(q_2)V(p)$ coupling is written as~\cite{Hagiwara:1986vm}
\begin{equation}
	\label{eq:vertex}
	\begin{aligned}
		\Gamma^{\alpha\beta\mu}_V(q_1,q_2,p) &= f_1^V(q_1-q_2)^\mu g^{\alpha\beta}-\frac{f_2^V}{m_W^2}(q_1-q_2)^\mu p^\alpha p^\beta +f_3^V(p^\alpha g^{\mu\beta}-p^\beta g^{\mu\alpha}) \\&+ if_4^V(p^\alpha g^{\mu\beta}+p^\beta g^{\mu\alpha})+if_5^V\epsilon^{\mu\alpha\beta\rho}(q_1-q_2)_\rho\\&-f_6^V\epsilon^{\mu\alpha\beta\rho}p_\rho - \frac{f_7^V}{m_W^2}(q_1-q_2)^\mu\epsilon^{\alpha\beta\rho\sigma}p_\rho (q_1-q_2)_\sigma,
	\end{aligned}
\end{equation}
with $f_i^V$ being the dimensionless functions of $p^2$. At the lowest order, the form factors are related to the anomalous parameters of Lagrangian given in Eq.~(\ref{eq:hag}) as~\cite{Degrande:2012wf}
\begin{equation}
	\begin{aligned}
		f_1^V &= g_1^V + \frac{s}{2m_W^2}\lambda_V,\\
		f_2^V &= \lambda_V,\\
		f_3^V &= g_1^V + \kappa_V +\lambda_V,\\
		f_i^V &= g_i^V \in i = 4,5,\\
		f_6^V &= \widetilde{\kappa}_V - \widetilde{\lambda}_V,\\
		f_7^V &= -\frac{\widetilde{\lambda}_V}{2}.
	\end{aligned}
\end{equation}
The above formalism was probed extensively at the LEP to test the predictions of SM with much success. The experiments were able to constrain the anomalous parameters to high accuracy with comparatively low energy and datasets. Nevertheless, the formalism suffers from subtle problems~\cite{Degrande:2012wf}; one can construct an infinite number of additional terms in Eq.~(\ref{eq:hag}) by adding derivatives. Each derivative would be accompanied by a factor of $m_W^{-1}$ since it is the only mass in the theory. These terms are not suppressed at higher energies, and thus, there is no principle to neglect them. These questions, along with the null results from experiments searching for new states, motivate the development of a model-independent framework to address the theory, either in whole or in part.

Up to now, the collider experiments have not observed any NP required to explain the shortcomings of the SM. In such scenarios, one assumes that the NP is too heavy to be produced at the CM energy of current colliders, and the effects of such states at the electroweak scale can be parametrized in terms of $SU(3)_C\times SU(2)_L\times U(1)_Y$ gauge invariant higher dimensional operators, that are suppressed by the NP scale. This framework of probing any NP deviation in observables considering higher dimensional operators is usually known as Standard Model effective field theory~(SMEFT) \cite{Buchmuller:1985jz,Grzadkowski:2010es}. The effect of NP set at some characteristic scale, $\Lambda$, is encoded in the Wilson coefficient associated with effective operators. The effective Lagrangian in the presence of higher-dimensional gauge and Lorentz invariant terms is given as
\begin{equation}
	\mathscr{L} = \mathscr{L}_{\text{SM}} + \sum_{i}\frac{C_i}{\Lambda^{d-4}}\mathscr{O}_i^{(d)},
\end{equation}
where $\mathscr{O}_i^{(d)}$s are the set of effective operators at dimension $d$ and $C_i$s are the associated Wilson coefficients of the corresponding operators. Assuming baryon and lepton number conservation, we consider the lowest order, {\it i.e.}, $d=6$, affecting $WWV$ couplings. In the HISZ basis, there are five $d=6$ operators that contribute to $WWV$ couplings. In terms of five $d=6$ operators, the effective Lagrangian becomes~\cite{Degrande:2012wf}
\begin{align}
	\begin{split}
		\delta\mathscr{L}^{d=6} =& \frac{C_B}{\Lambda^2}(D_{\mu}\Phi)^{\dagger}B^{\mu \nu}(D_{\nu}\Phi) + \frac{C_W}{\Lambda^2}(D_{\mu}\Phi)^{\dagger}W^{\mu \nu}(D_{\nu}\Phi) + \frac{C_{WWW}}{\Lambda^2}\text{Tr}[W_{\mu \nu} W^{\nu \rho} W_{\rho}^{\mu}]\\
		&+ \frac{C_{\widetilde{WWW}}}{\Lambda^2}\text{Tr}[\widetilde{W}_{\mu \nu} W^{\nu \rho} W_{\rho}^{\mu}] + \frac{C_{\widetilde{W}}}{\Lambda^2}(D_{\mu}\Phi)^{\dagger}\widetilde{W}^{\mu \nu}(D_{\nu}\Phi),
	\end{split}
	\label{eq:dim6.ops}
\end{align}
where $C_i\in\{C_{WWW},C_W,C_B\}$ are three CP-even and $C_{\widetilde{i}}\in\{C_{\widetilde{W}},C_{\widetilde{WWW}}\}$ are two CP-odd effective couplings. These couplings encode the effect of any BSM physics at the electroweak scale. Here, $\Phi$ is the SM Higgs doublet, $D_{\mu}$ is the covariant derivative, $B_{\mu \nu}$ and $W_{\mu \nu}$ are the $U(1)$ and $SU(2)$ field strength tensors defined as
\begin{equation*}
	\begin{aligned}
		D_\mu &= \partial_\mu + i\frac{g}{2}\tau^i W^i_\mu + i\frac{g^\prime}{2}B_\mu,\\
		W_{\mu\nu} &= i\frac{g}{2}\tau^i\left(\partial_\mu W^i_\nu -\partial_\nu W^i_\mu + g\epsilon_{ijk}W^j_\mu W^k_\nu\right),\\
		B_{\mu\nu} &= i\frac{g^\prime}{2}\left(\partial_\mu B_\nu - \partial_\nu B_\mu\right),
	\end{aligned}
\end{equation*}
with $g$ and $g^\prime$ are $SU(2)$ and $U(1)$ couplings constant, respectively. The anomalous part of Eq.~(\ref{eq:hag}) can be written in terms of effective couplings as~\cite{Degrande:2012wf}
\begin{equation}
	\centering
	\label{eqn:relation}
	\begin{aligned}
		&\Delta g^Z_1=C_W \frac{m_Z^2}{2\Lambda^2}, \\ 
		&\Delta g_1^\gamma = g_4^{V}=g_5^{V}=0,\\
		&\lambda^{\gamma}=\lambda^{Z}=C_{WWW}\frac{3g^2m_W^2}{2\Lambda^2},\\
		&\Delta{\kappa}^{\gamma}=(C_W+C_B)\frac{m_W^2}{2\Lambda^2},\\
		&\Delta{\kappa}^{Z}=(C_W-C_B \tan^2\theta_W)\frac{m_W^2}{2\Lambda^2},\\
		&\widetilde{\kappa}^{\gamma}=C_{\widetilde{W}}\frac{m_W^2}{2\Lambda^2},\\
		&\widetilde{\kappa}^{Z}=-C_{\widetilde{W}} \tan^2\theta_W\frac{m_W^2}{2\Lambda^2},\\
		&\widetilde{\lambda}^{\gamma}=\widetilde{\lambda}^{Z}=C_{\widetilde{WWW}}\frac{3g^2m_W^2}{2\Lambda^2}.
	\end{aligned}
\end{equation}
Thus, from the above equation, it becomes clear that five dimension-6 operators become a complete set to describe the effective Lagrangian describing general $WWV$ couplings. Also, the number of independent anomalous couplings can be reduced by writing down two more relations from above Eq.~\eqref{eqn:relation}
\begin{equation}
	\Delta g^Z_1=\Delta \kappa^Z + \tan^2 \theta_W \Delta \kappa^{\gamma}, \qquad \tilde{\kappa}^{Z} + \tan^2 \theta_W \tilde{\kappa}^{\gamma}=0. 
\end{equation}

\noindent
Search for NP through dimension-6 cTGCs has been performed via different di-boson productions at the CMS experiment in the LHC. The most stringent limit on each dimension-6
cTCGs at 95\% CL is listed in Table \ref{tab:95cl.limit}. In the subsequent analysis, we will determine the optimal sensitivity of dimension-6 cTGCs through $WW$ production. We focus our analysis on the CLIC running with $\sqrt{s}=3$ TeV, $\mathfrak{L}_{\text{int}}=1000~\text{fb}^{-1}$, and polarized electron beam.

\begin{table}[htb!]
	\centering
	{\renewcommand{\arraystretch}{1.3}%
		\begin{tabular}{ |c|c|c| } 
			\hline
			NP couplings & Limits ({$\rm{TeV^{-2}}$}) & References \\ 
			\hline
			$C_{B}/\Lambda^2$ & [-8.78, +8.54] & CMS \cite{CMS:2019ppl}\\
			$C_{W}/\Lambda^2$ & [-3.10, +0.30] & CMS \cite{CMS:2021icx}\\ 
			$C_{WWW}/\Lambda^2$ & [-0.90, +0.91] & CMS \cite{CMS:2021foa} \\ 
			$C_{\widetilde{WWW}}/\Lambda^2$ & [-0.45, +0.45] & CMS \cite{CMS:2021foa}\\ 
			$C_{\widetilde{W}}/\Lambda^2$ & [-20, +20] & CMS \cite{CMS:2021foa}\\
			\hline
	\end{tabular}}
	\caption{Experimental constraints (95\% CL) on dimension-6 cTGCs from the CMS experiments at the LHC.}
	\label{tab:95cl.limit}
\end{table}

\section{$WW$ production at the $e^+e^-$ colliders}
\label{sec:ww.prod}
\begin{figure}[t]
	$$
	\includegraphics[height=4.3cm,width=7cm]{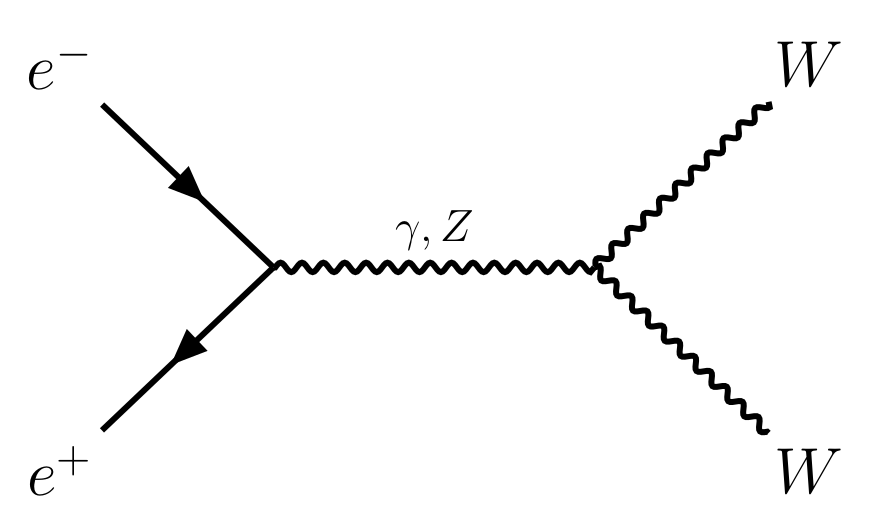}
	\includegraphics[height=4.3cm,width=7cm]{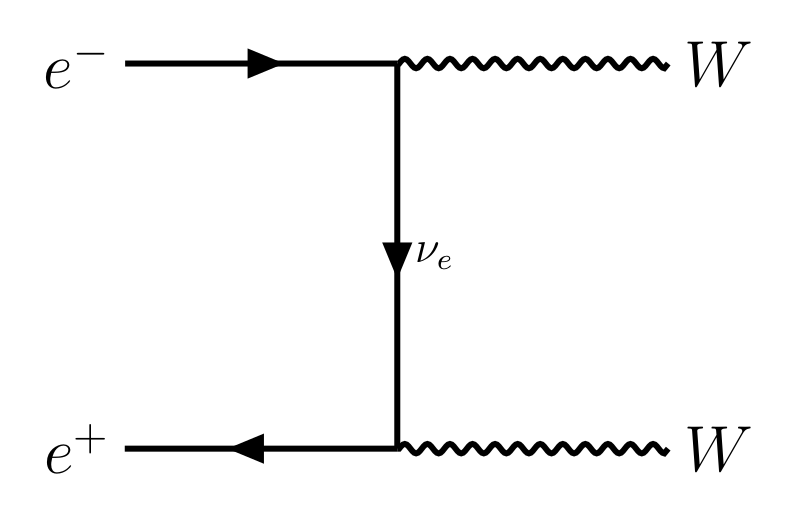}
	$$
	$$
	\includegraphics[height=4.3cm,width=7cm]{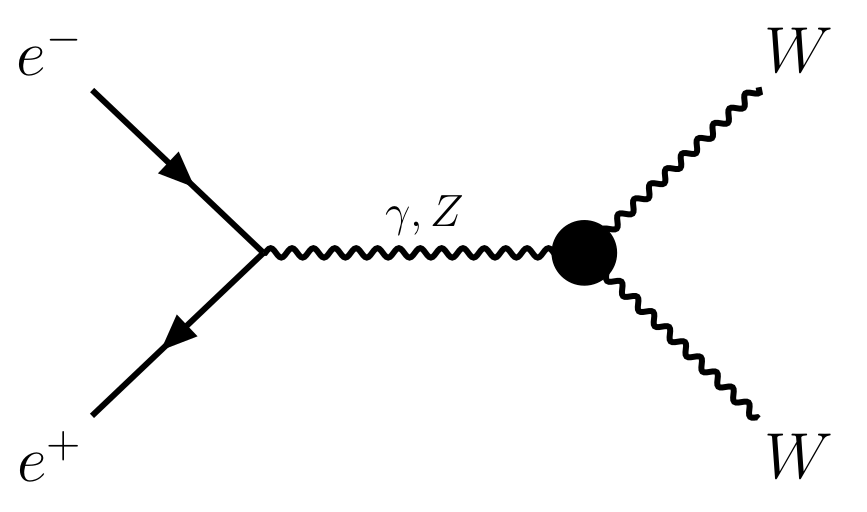}
	$$
	\caption{Schematic Feynman diagrams for $WW$ production at the $e^+e^-$ colliders. Top panel: SM contribution, bottom: BSM contribution is highlighted in the dark blob appearing through the cTGCs vertex.}
	\label{fig:feyn.diag}
\end{figure}

\noindent
In $e^+e^-$ colliders, at tree level, $W$-boson pair production is predominantly driven by $s$-channel process mediated by $\gamma$ and $Z$ bosons, along with a $t$-channel contribution mediated by neutrinos within the SM (top panel of Fig.~\ref{fig:feyn.diag}). For the BSM scenario considered here, dimension-6 operators contribute to $WW$ production through $\gamma$- and $Z$-mediated $s$-channel mediation\footnote{The dimension-6 cTGCs can also be explored through vector boson fusion (VBF) processes at the high-energy $e^+e^-$ colliders.} shown in the bottom of Fig.~\ref{fig:feyn.diag}. 
The quantum state of the $W^+W^-$ system produced at the $e^+e^-$ collisions can be fully described by the helicity density matrix
\begin{equation}
  \rho(\lambda_{W^-}, \lambda_{W^+}; \lambda'_{W^-}, \lambda'_{W^+}) \propto M(\lambda_{W^-}, \lambda_{W^+})\, M^*(\lambda'_{W^-}, \lambda'_{W^+}),
\end{equation}
where $\lambda_{W^\pm}, \lambda'_{W^\pm} \in \{-1, 0, +1\}$ denote the helicities of the $W^\pm$ bosons in the amplitude and its complex conjugate, respectively. The proportionality factor includes the flux and phase-space terms. This $9\times 9$ matrix encapsulates all spin correlations and quantum interference effects in the diboson system.

To analyze its structure, we classify the helicity configurations into three distinct combinations. The longitudinal-longitudinal (LL) combination corresponds to $\lambda_{W^-} = \lambda_{W^+} = \lambda'_{W^-} = \lambda'_{W^+} = 0$. The transverse-transverse (TT) cobination contains elements with $\lambda_{W^\pm}, \lambda'_{W^\pm} \in \{\pm1\}$, and the longitudinal-transverse (LT) combination refers to all remaining configurations where one of the $W$ bosons carries helicity zero and the other has helicity $\pm1$, in either the amplitude or its conjugate. When helicities are summed inclusively without resolution, we refer to the result as the unresolved case.

The individual helicity amplitudes that enter the construction of $\rho$ can be written as~\cite{Hagiwara:1986vm}
\begin{equation}
  M_{\sigma,\bar{\sigma};\lambda,\lambda^\prime} = \sqrt{2}e^2\, \widetilde{M}_{\sigma,\bar{\sigma};\lambda,\lambda^\prime}(\Theta)\, d^{\max(|\Delta\sigma|,|\Delta\lambda|)}_{\Delta\sigma,\Delta\lambda}(\theta),
  \label{eq:amp.ww}
\end{equation}
where $\sigma$ and $\bar{\sigma}~(=\pm1)$ are the helicities of the incoming electron and positron, while $\lambda$, $\lambda^\prime$ are the helicities of the outgoing $W^-$ and $W^+$. The angle $\theta$ denotes the scattering angle of the $W^-$ relative to the electron direction in the center-of-mass frame, and $d^j_{mm'}(\theta)$ are the Wigner $d$-functions, defined in Appendix~\ref{sec:wig.func}. Here, $\Delta\sigma = (\sigma - \bar{\sigma})/2$ and $\Delta\lambda = \lambda - \lambda'$. In the high-energy limit $\sqrt{s} \gg m_e$, angular momentum conservation implies $\bar{\sigma} = -\sigma$, so only $\Delta\sigma = \pm1$ amplitudes contribute. In this limit, the reduced helicity amplitude becomes~\cite{Hagiwara:1986vm}
\begin{equation}
  \begin{aligned}
    \widetilde{M}_{\sigma,-\sigma;\lambda,\lambda^\prime} &= \frac{\beta}{\sin^2\theta_W} \left( -\frac{1}{2}\delta_{\sigma,-1} + \sin^2\theta_W \right) A^Z_{\lambda,\lambda^\prime} \frac{s}{s - m_Z^2} - \beta A^\gamma_{\lambda,\lambda^\prime} \\
    &\quad + \delta_{\sigma,-1} \frac{1}{2\beta\sin^2\theta_W} \left[ B_{\lambda,\lambda^\prime} - \frac{C_{\lambda,\lambda^\prime}}{1 + \beta^2 - 2\beta \cos\theta} \right],
  \end{aligned}
\end{equation}
where $\beta^2 = 1 - \gamma^{-2}$ with $\gamma = \sqrt{s}/(2 m_W)$, and $\theta_W$ is the weak mixing angle. The terms $A^\gamma$ and $A^Z$ correspond to $s$-channel photon and $Z$ boson exchange, while $B$ and $C$ arise from $t$-channel neutrino exchange.

The CP-violating dimension-6 operators, introduced in Eq.~\eqref{eq:vertex}, contribute only to helicity configurations satisfying $\lambda + \lambda' \neq 0$~\cite{Chang:1993vv}, and therefore do not affect the LL combination. Moreover, the contribution of the $C_B/\Lambda^2$ operator to the TT sector vanishes due to Ward identities. These helicity selection rules are summarized in Table~\ref{tab:BPs}, where a check mark indicates that a given operator contributes to at least one helicity configuration within the corresponding combination.

\begin{table}[!h]
	\centering
	{\renewcommand{\arraystretch}{1.1}%
		\begin{tabular*}{1\textwidth}{@{\extracolsep{\fill}}ccccccc@{}}
			\hline
			\multirow{2}*{Coupings} &
			\multicolumn{4}{c}{Helicity combinations} \\
			\cline{2-5}
			& LL $(00)$ & LT+TL $(0+/-)$ & TT $(++,+-,--)$ & Unresolved \\
			\hline
			$C_B/\Lambda^2$ & \ding{51} & \ding{51} & \ding{55} & \ding{51}  \\
			$C_W/\Lambda^2$  & \ding{51} & \ding{51} &\ding{51}& \ding{51}  \\
			$C_{WWW}/\Lambda^2$  & \ding{51} & \ding{51} &\ding{51}& \ding{51} \\
			$C_{\widetilde{WWW}}/\Lambda^2$  & \ding{55} &\ding{51}&\ding{51} & \ding{51} \\
			$C_{\widetilde{W}}/\Lambda^2$  & \ding{55} &  \ding{51} &\ding{51}& \ding{51} \\
			\hline
	\end{tabular*}}
	\caption{Contributions of different NP couplings to the total cross-section for different helicity combinations. \ding{51} (\ding{55}) indicates that particular couplings are present (absent) for a specific helicity combination.}
	\label{tab:BPs}
\end{table}
Further, the upcoming $e^-e^+$ collider viz., CLIC will be colliding an initially polarized beams, which would change the overall differential rate. The differential cross-section for partially polarized initial beams $\left(-1\le P_{e^{\pm}}\le+1\right)$ can be expressed as
\begin{align}
	\begin{split}
		\frac{d\sigma(P_{e^+},\,P_{e^-})}{d\Omega} =&   \frac{(1-P_{e^-})(1+P_{e^+})}4 \left( \frac{d\sigma}{d\Omega}\right)_{LR} +\frac{(1+P_{e^-})(1-P_{e^+})}4 \left( \frac{d\sigma}{d\Omega}\right)_{RL}.
		\label{eq:diff.cs}
	\end{split}
\end{align}	
Here, $\left( d\sigma/d\Omega \right)_{ij}$ represents the differential cross-sections for $WW$ production with helicities $ i \in \{L, R\} $ and $ j \in \{L, R\} $, where $ L $ and $ R $ denote left- and right-handed helicities of the initial beams, respectively.
\begin{figure}[t]
	\centering
	\includegraphics[height=5cm,width=4.95cm]{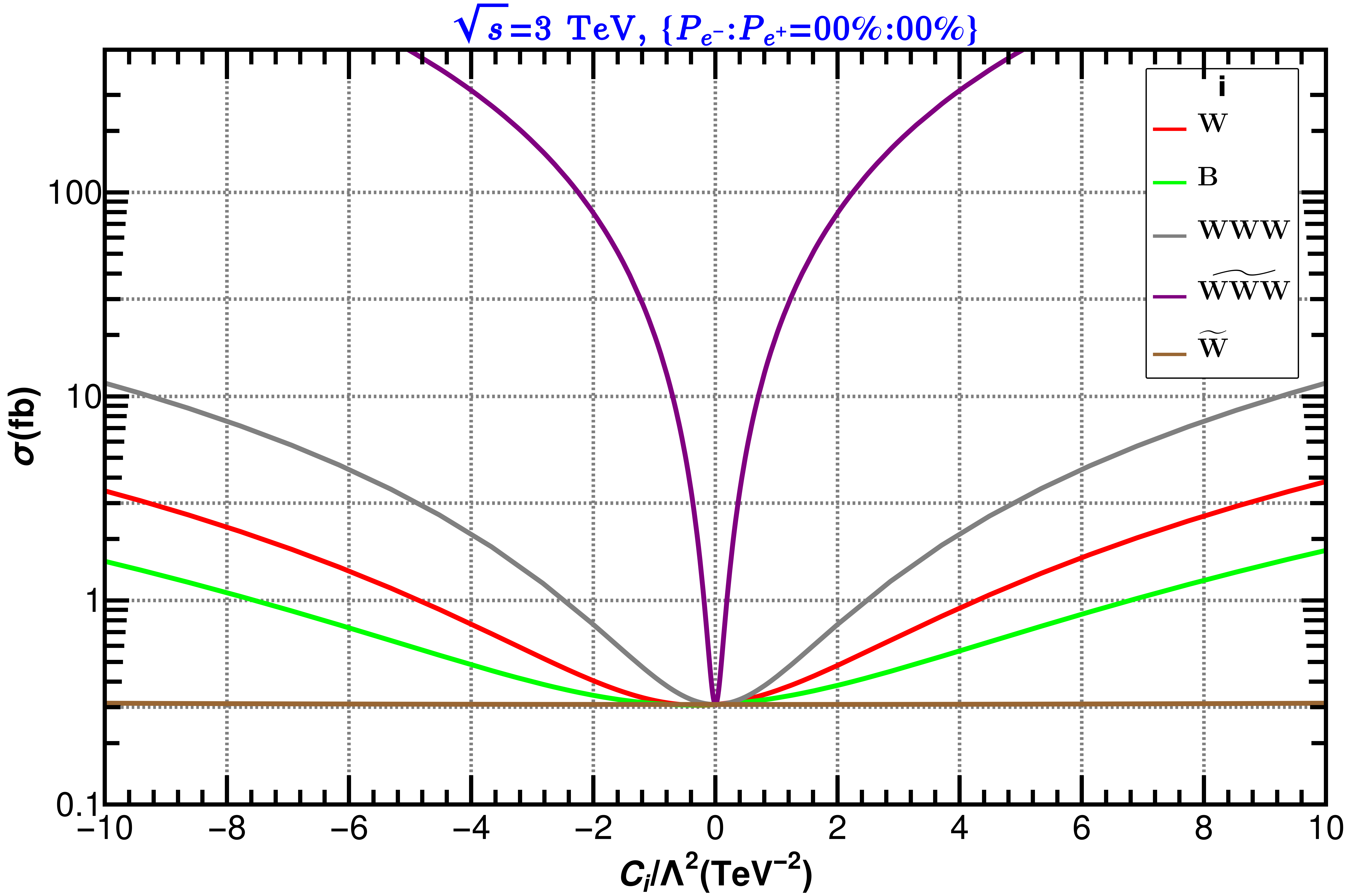}
	\includegraphics[height=5cm,width=4.95cm]{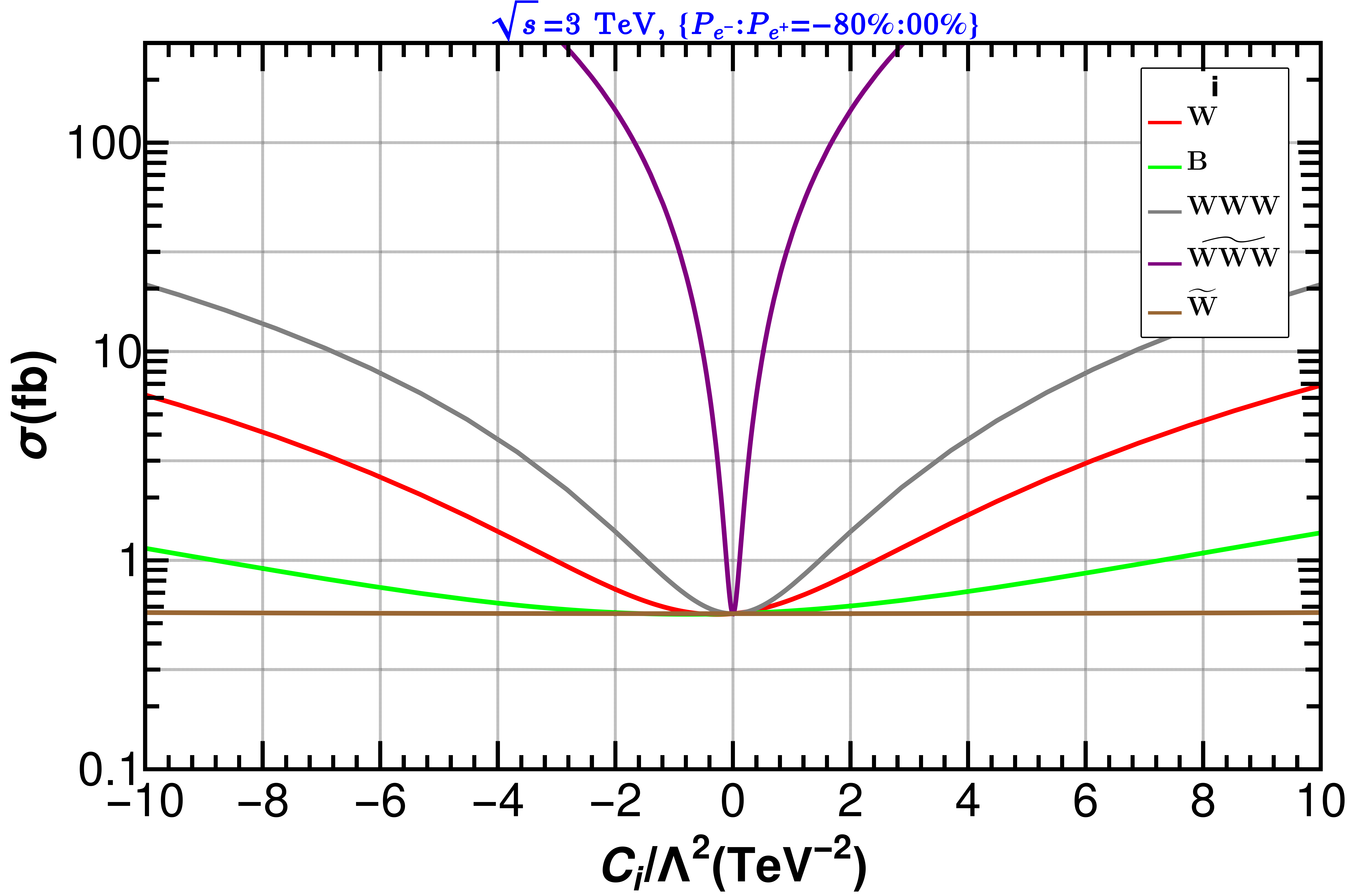}
	\includegraphics[height=5cm,width=4.95cm]{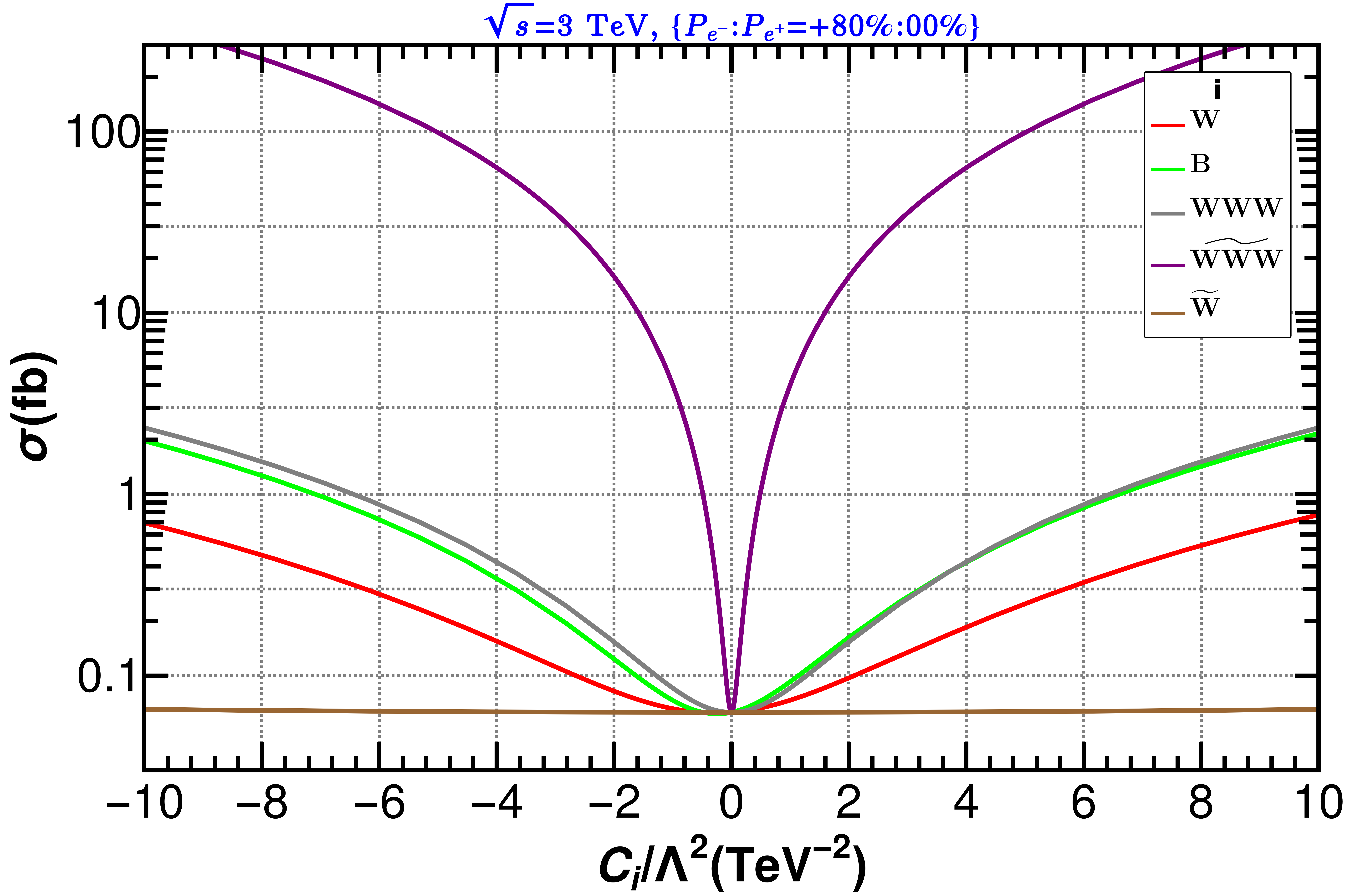}
	\caption{Variation of total cross-section with different dimension-6 effective couplings at $\sqrt{s}=3$ TeV CLIC. Left: Unpolarized beam; middle:$\{P_{e^-}:P_{e^+}=-80\%:00\%\}$, right: $\{P_{e^-}:P_{e^+}=+80\%:00\%\}$. Here we consider the unresolved helicity $W$ boson pair.}
	\label{fig:xsec}
\end{figure}
Dominant CP-even BSM contributions to the $WW$ cross-section originate from the interference term between the SM and dim-6 operators when $|C_i/\Lambda^2|<1~\text{TeV}^{-2}$. In this region, the variation of total cross-section with the effective coupling is linear. Pure BSM contribution starts to dominate when $|C_i/\Lambda^2|>1~\text{TeV}^{-2}$. For the CP-odd scenario, there is no interference between SM and BSM. Therefore, the BSM contribution is directly proportional to $C^2_{i}/\Lambda^4$ for the whole range of the effective coupling. As a result, dimension-8 SMEFT contribution may be comparable to the $C^2_{\widetilde{WWW}}/\Lambda^4$ term. For simplicity, we neglect the effect of dimension-8 contribution and stick to the dimension-6 SMEFT. In Fig.~\ref{fig:xsec}, we show the variation total cross-section with the dimension-6 effective couplings for three different polarization combinations considering unresolved helicity of $WW$ production. $C_{\widetilde{WWW}}/\Lambda^2$ has the most dominant contribution to the $WW$ production while $C_{\widetilde{W}}/\Lambda^2$ has the least. 

The reconstruction of $W$ boson helicity at the collider experiments cannot be achieved with complete accuracy. The LEP experiment measured 
$W$ boson helicities with approximately 83\% accuracy \cite{L3:2002ygr}. At the LHC, this measurement can achieve above 90\% accuracy \cite{ATLAS:2019bsc}. Furthermore, using deep neural networks (DNN), it has been demonstrated in \cite{Kim:2021gtv} that 
$W$ boson helicities can be measured with approximately 98\% accuracy at the LHC. These determination are subject to the different colliders observables and different feducial regions. However, since our focus is not on the helicity measurement itself but rather on its impact on dimension-6 cTGCs, we take a conservative estimate of 90\% accuracy in $W$ boson helicity measurement in our phenomenological analysis. A 10\% uncertainty in the $W$ boson polarization measurement results in an approximate 6\% reduction in the sensitivity to NP couplings. In the next section, we consider the semi-leptonic process $WW \to q q' \ell \nu_{\ell}$ as our final state signal, which is a very good optimization between event rate as well as a relatively clean final state at the $e^+e^-$ colliders. 

\section{Collider analysis}
\label{sec:col}
\begin{figure}[t]
	\centering
        \includegraphics[width=0.49\textwidth]{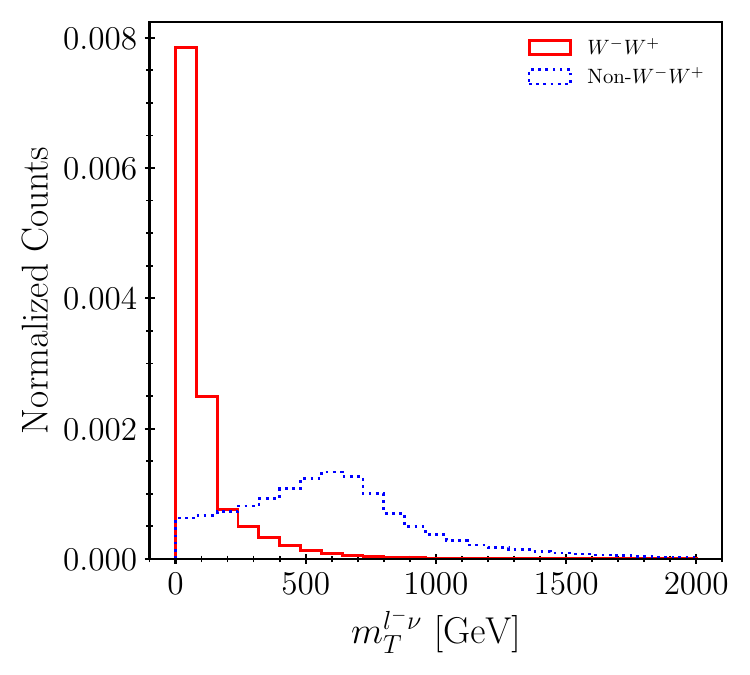}
	\includegraphics[width=0.49\textwidth]{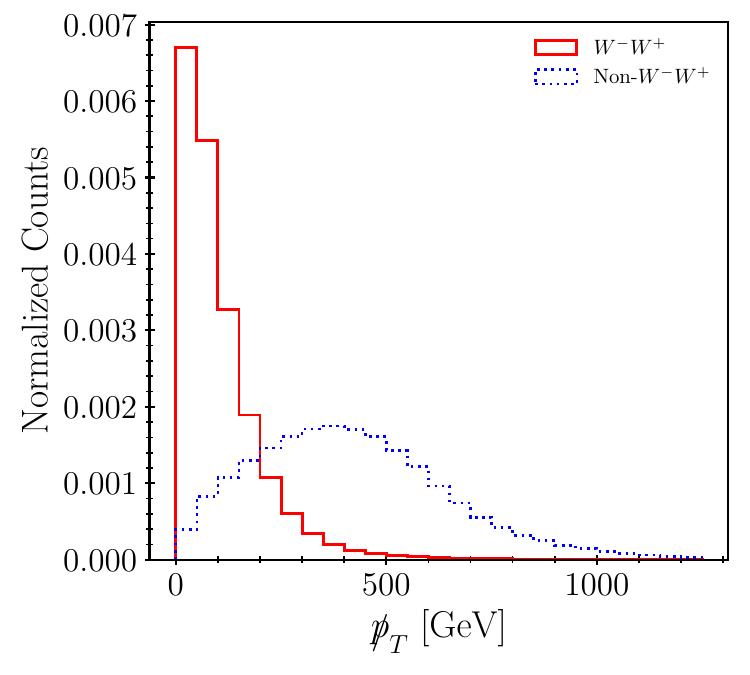}\\
	\includegraphics[width=0.49\textwidth]{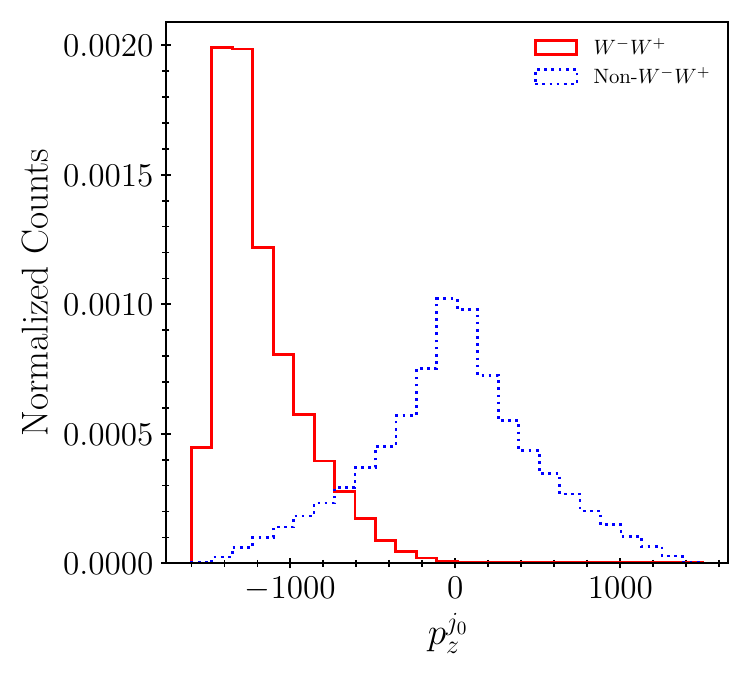}
	\includegraphics[width=0.49\textwidth]{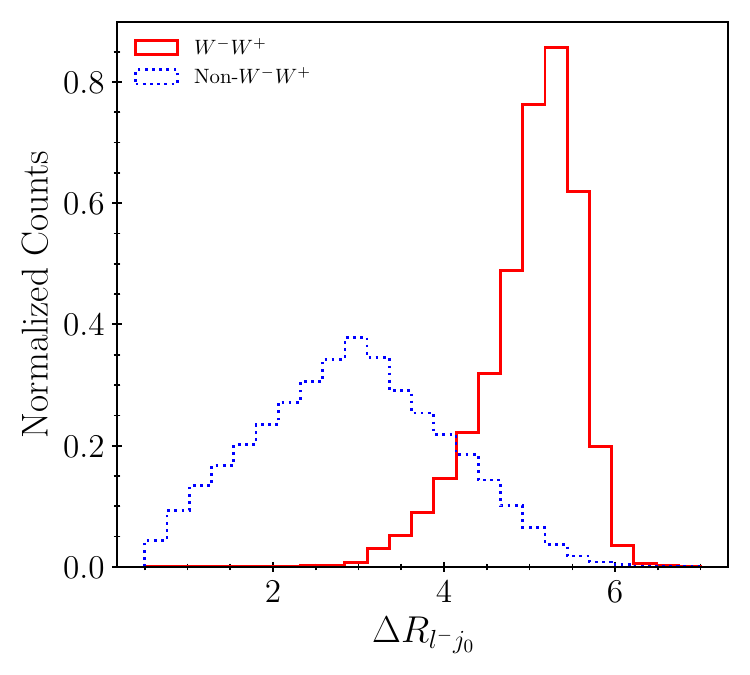}
	\caption{Normalized distributions for different collider observables at the detector level at $\sqrt{s}=3$ TeV CLIC. Top left: transverse mass of lepton and missing neutrino ($m_T^{l^-\nu}$), top right: missing transverse momenta ($p_T^\nu$); bottom left: longitudinal momenta of leading jet ($p_z^{j_0}$), bottom right: angular separation between the leading jet and negatively charged lepton ($\Delta R_{l^-j_0}$).  }
	\label{fig:kindelphes}
\end{figure}

\begin{figure}[!htb]
	\centering
	\includegraphics[width=0.49\textwidth]{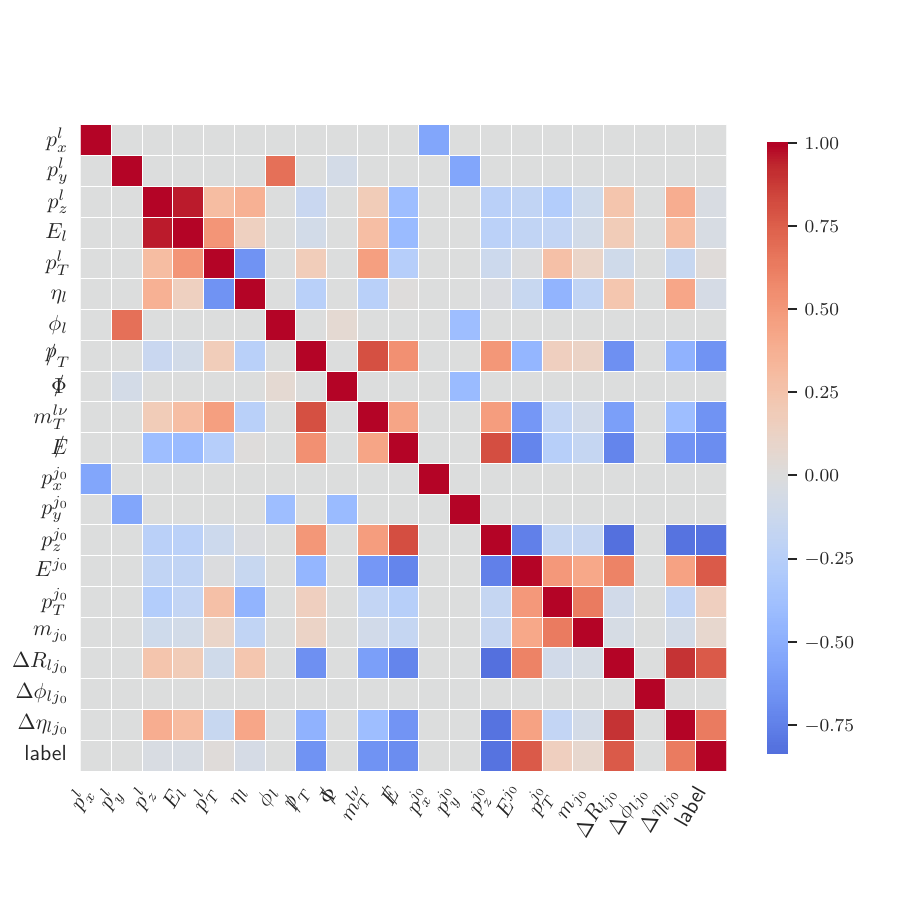}
  \includegraphics[width=0.49\textwidth,height=0.28\textheight]{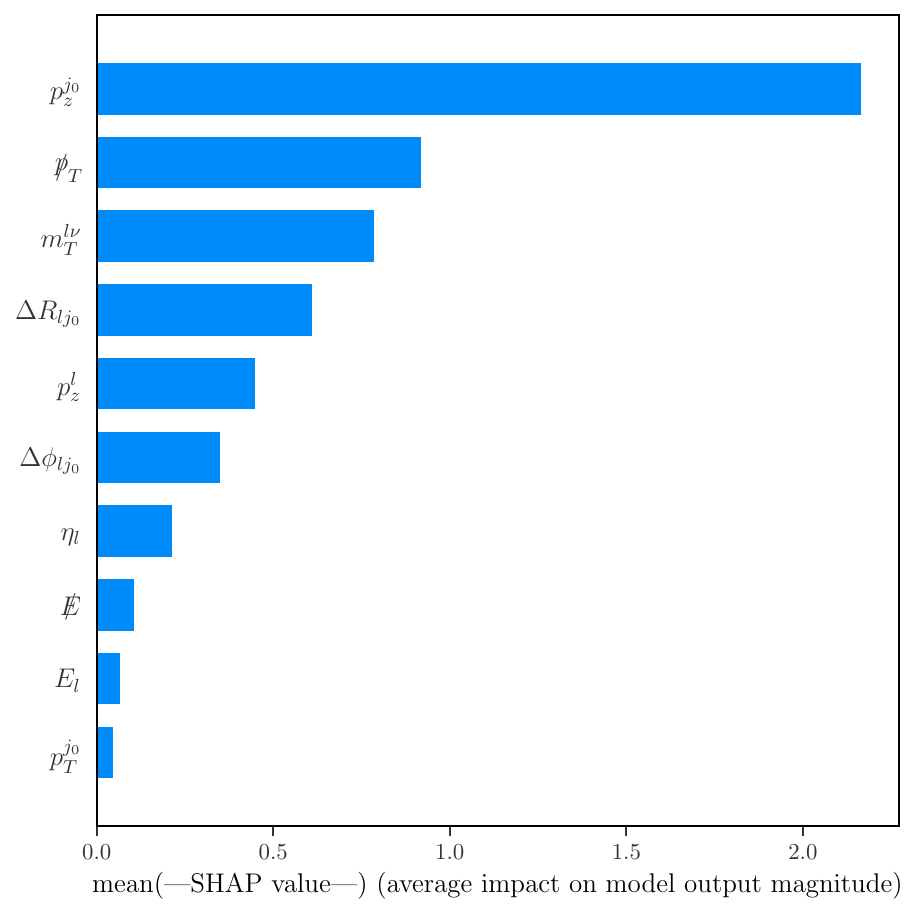}
  \includegraphics[width=0.49\textwidth]{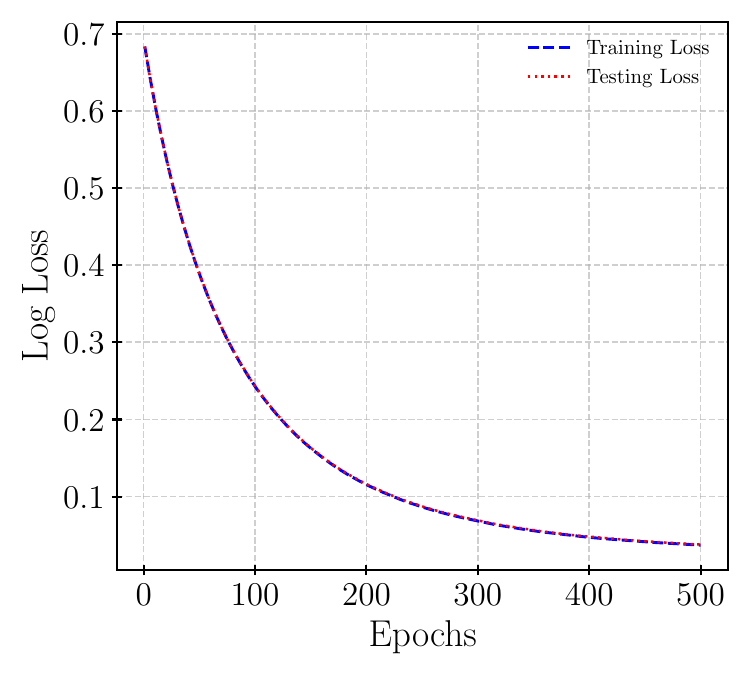}
  \includegraphics[width=0.49\textwidth]{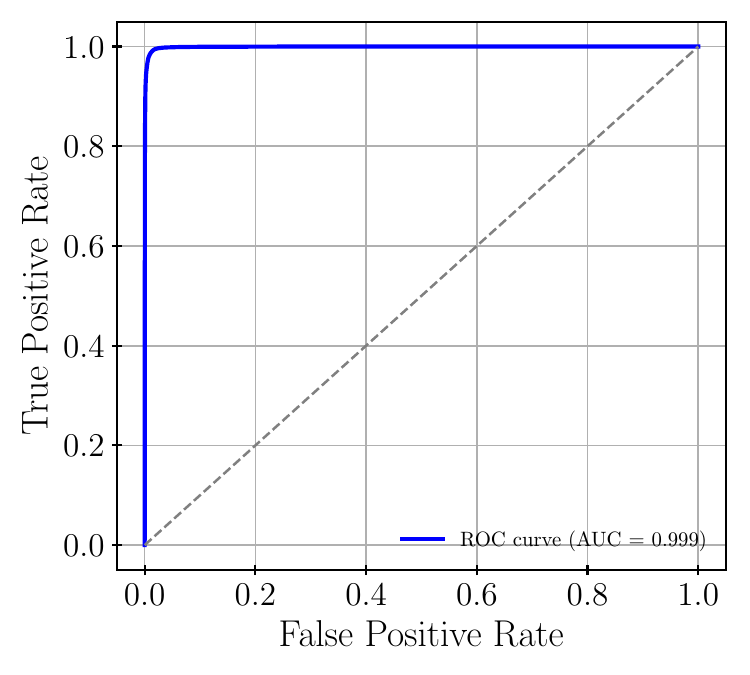}
	\caption{In the top row, we show the Pearson correlation (left panel) among features and class of events and ten features with highest Shapely additive explanations (SHAP) values. In the bottom row, we show the training and testing loss (left) and ROC curve (right) for the BDT model. The loss curves show stable convergence with no significant deviation, while the high AUC of ~$99\%$ indicates excellent signal-background separation, making the semileptonic WW decay nearly background-free.}
	\label{fig:roc}
\end{figure}

The major background for the semi-leptonic decay of $W^-W^+$ resonant amplitudes arises from continuum processes with one non-resonant $W$ boson. To analyze signal and background events, we generate a set of Monte Carlo events using {\tt MadGraph5$\_$aMC$@$NLO} \cite{Alwall:2011uj}, with the model implemented in {\tt FeynRules} \cite{Alloul:2013bka}. The showering of colored partons and hadronization of final states is done using {\tt Pythia} \cite{Sjostrand:2014zea}. Finally, the detector simulation is implemented with {\tt Delphes} \cite{deFavereau:2013fsa}. The generation of signal events is direct, while for the generation of continuum backgrounds, we used the diagram filter plugin within MG5, removing all $WW$ di-resonant amplitudes. The events are generated at $\sqrt{s}=3$ TeV CLIC with following parton level kinematic cuts,
\begin{equation}
	\begin{aligned}
		&p_T^j \ge 20~\mathrm{GeV},~~p_T^l \ge 10~\mathrm{GeV},~~|\eta_j| \le 5,\\&|\eta_l| \le 2.5,~~ |\Delta R_{jj}| \ge 0.4,~~|\Delta R_{lj}| \ge 0.4,
	\end{aligned}
\end{equation}
where $\Delta R_{ab} = \sqrt{(\eta_a-\eta_b)^2 + (\phi_a-\phi_b)^2}$ is the geometric distance between $a$ and $b$ particles. Here, we consider the case where the $W^+$ decays hadronically and the $W^-$ decays leptonically\footnote {The reversed decay configuration has also been included in our calculations to estimate the sensitivity in the dimension-6 CTGCs.}. We show the normalized event distributions of transverse mass of lepton and missing neutrino ($m_T^{l^-\nu}$), missing transverse momenta ($\slashed{p}_T$), longitudinal momenta of the leading jet ($p_z^{j_0}$) and $\Delta R$ between the negatively charged lepton and hardest jet at the detector level in Fig.~\ref{fig:kindelphes}. It is also to be noted that due to large $\sqrt{s}$, the jets decayed from the $W^+$ boson will be highly boosted and can lead to a single collimated jet and the analysis of such fat jets could potentially assist in signal separation from backgrounds. However, in our current study, we do not perform any jet substructure analysis as most of the preliminary kinematic variables show significant differences in the two kinds of events. In the case of signal events, $m_T^{l^-\nu}$ distribution shows a peak around the pole mass of $W$ boson, while for the background, we notice rather a broad distribution with shifted peak around 600 GeV. A similar distinction is evident in the case of $\slashed{p}_T$. The longitudinal momenta of the leading jet also prove to be a significant observable to suppress background events. For the signal events, the $p_Z$ shows a skew distribution with a peak around $1.2$ TeV, while for the background, the distribution is symmetric around zero. Finally, we noticed a large angular separation $\Delta R \in [0.5,6.0]$ between the charged lepton and hardest jet in the case of background, while for the signal events, the angular separation remains fairly tight with $\Delta R \in [3.0,6.0]$. One can employ these distributions in order to reduce the background contribution to increase the overall signal significance. The goodness of these observables in order to select signal events is highlighted in the Pearson correlation plot shown in left panel (top row) of Fig.~\ref{fig:roc}. The Figure suggests that any one of the observable $\mathcal{O} \in \{p_z^{j_0}, E_\nu,m_T^{l^-\nu},\Delta R_{l^-j_0}\}$ could distinguish the two class of events with an accuracy in the range of $50 - 75\%$.

With the rise of modern machine learning tools, multivariate analysis has become a standard approach in collider studies. Among various machine learning techniques, boosted decision tree (BDT) algorithms \cite{Cornell:2021gut,Coadou:2022nsh} are particularly popular due to their robustness and high accuracy, even when dealing with smaller datasets. In our analysis, we develop a simple BDT model to maximize (over cut-based analysis) the signal vs. background classification. The features used in the BDT model are listed in Table~\ref{tab:bdtfeat}.

Our BDT model is fairly simple and is implemented with {\tt XGBoost} \cite{Chen:2016btl} for binary classification of signal and background events. It consists of an ensemble of depth-4 decision trees, trained with 500 boosting rounds and a learning rate of 0.01, using binary logistic loss as the objective. The input features, consisting of 20 kinematic variables, are standardized using $z$-score normalization before training. The dataset of size $1.5\times 10^6$ events are taken out of which $70\%$ are used for training and remaining datasets are reserved for testing. The model performance is monitored using log-loss. In Fig.~\ref{fig:roc}, we present the training and testing loss curves, noting no significant deviation between them, indicating the absence of overfitting. Additionally, the ROC curve, also shown in Fig.~\ref{fig:roc}, exhibits an area under the curve (AUC) of approximately $99\%$, demonstrating excellent classification performance. This high AUC implies that the background can be efficiently removed, rendering the leading order semileptonic decay of $W^-W^+$ an almost background-free process. Hence, the rest of the analysis does not take into account of the background contamination in probe of anomalous couplings.

\begin{table}[H]
	\centering
	\begin{tabular}{cccc}\\
		\hline
		Features & Description & Features & Description\\
		\hline
		$p_{l^-/j_0}$ & 4-momenta of lepton/hardest jet   & $p_T^{l^-/j_0}$ & Transverse momenta   \\
		$\eta_{l^-/j_0}$ & Pseudorapidity & $\phi_{l^-/j_0}$ & Azimuth orientation \\
		$\Delta R_{l^-j^0}$ & Angular distance between $l^-$ and $j_0$& $\Delta\eta_{l^-j^0}$ & Separation of pseudorapidity\\
		$\Delta\phi_{l^-j^0}$ & Azimuth separation & $m_{j^0}$ & Mass of hardest jet \\
		$p_T^\nu$ & Missing transverse momenta & $E_\nu$ & Missing energy\\
		$m_T^{l^-\nu}$ & Transverse mass of $l^-$ and $\nu$ & $\phi^\nu$ & Azimuth of neutrino \\
		\hline
	\end{tabular}
		\caption{List of features used in the BDT model to perform a signal vs. background classification.}
	\label{tab:bdtfeat}
\end{table}

\section{Optimal Observable Technique}
\label{sec:oot}
The Optimal Observable Technique (OOT) is a powerful tool for optimally determining the statistical sensitivity of any NP coupling. In this section, we provide a brief overview of the mathematical framework underlying OOT, which has been thoroughly detailed in previous studies \cite{Diehl:1993br,Gunion:1996vv,Bhattacharya:2021ltd}. Typically, any observable, {{\it e.g.,} differential cross-section} that incorporates contributions from both the SM and BSM can be represented as
\begin{equation}
	\mathcal{O}(\phi)=\frac{d\sigma}{d\phi}=g_i f_i(\phi),
\end{equation}
Here, $g_i$s represent functions of the NP coefficients, while $f_i(\phi)$ depends on the phase-space variable $\phi$. Our analysis focuses on the process $e^+e^- \to  W^+W^-$, therefore, the phase-space variable of interest is the cosine of the angle of the outgoing particle ($\cos \theta$) in the CM frame. Depending on the specific observable or process under study, other variables can be altered in place of $\phi$.

Our objective is to estimate $g_i$, which can be accomplished by applying a suitable weighting function, $w_i(\phi)$:
\begin{equation}
	g_i=\int w_i(\phi) \mathcal{O}(\phi) d\phi,
\end{equation}
While various options for $w_i(\phi)$ are possible, there is a specific choice that optimizes the covariance matrix, $V_{ij}$, minimizing statistical uncertainties in NP couplings. For this optimal choice,  $V_{ij}$  is given by
\begin{equation}
	V_{ij} \propto \int w_i(\phi) w_j(\phi) \mathcal{O}(\phi) d\phi,
\end{equation}
Consequently, the weighting functions that meet the optimal condition $\delta V_{ij} = 0$ are
\begin{equation}
	w_i(\phi)=\frac{M^{-1}_{ij}f_j(\phi)}{\mathcal{O}(\phi)},
\end{equation}
where
\begin{equation}
	M_{ij}=\int \frac{f_i(\phi) f_j(\phi)}{\mathcal{O}(\phi)}d\phi.
\end{equation}
The optimal covariance matrix is formulated as follows:
\begin{equation}
	V_{ij}=\frac{M^{-1}_{ij}}{\mathfrak{L}_{\tt int}}.
\end{equation}
Here, \(\sigma_T = \int \mathcal{O}(\phi) \, d\phi\). $N$ signifies the total number of events which is expressed as $N = \sigma_T \mathfrak{L}_{\text{int}}$.

The function $\chi^2$, which determines the optimal constraint for NP couplings, is defined as
\begin{equation}
	\chi^2=\sum_{ij}(g_i-g_i^0)(g_j-g_j^0)V^{-1}_{ij},
	\label{eq:oot.chi.sqr}
\end{equation}
where $ g_0 $ represents `seed values' that depend on the specific NP scenario. The limit defined by $\chi^2 \leq n^2$ corresponds to $n\sigma$ standard deviations from these seed values $g_0$, setting the optimal constraint for NP couplings under the fact that the covariance matrix $V_{ij}$ is minimized. Using the definition of the $\chi^2$ function provided in Eq.~\eqref{eq:oot.chi.sqr}, the optimal constraints on NP couplings are explored in the following section.

\begin{figure}[t]
	$$   
	\includegraphics[height=5cm,width=4.95cm]{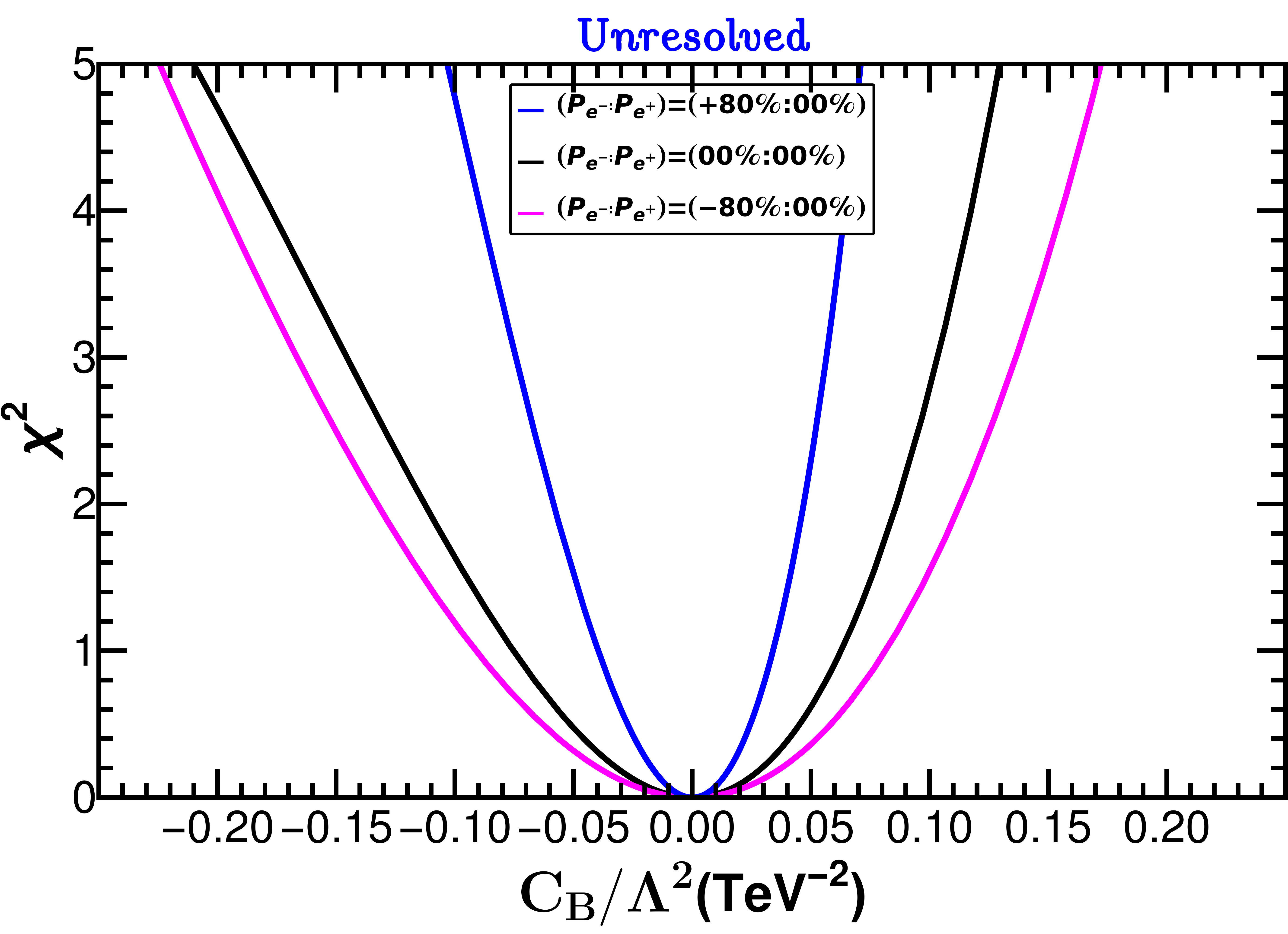}~
	\includegraphics[height=5cm,width=4.95cm]{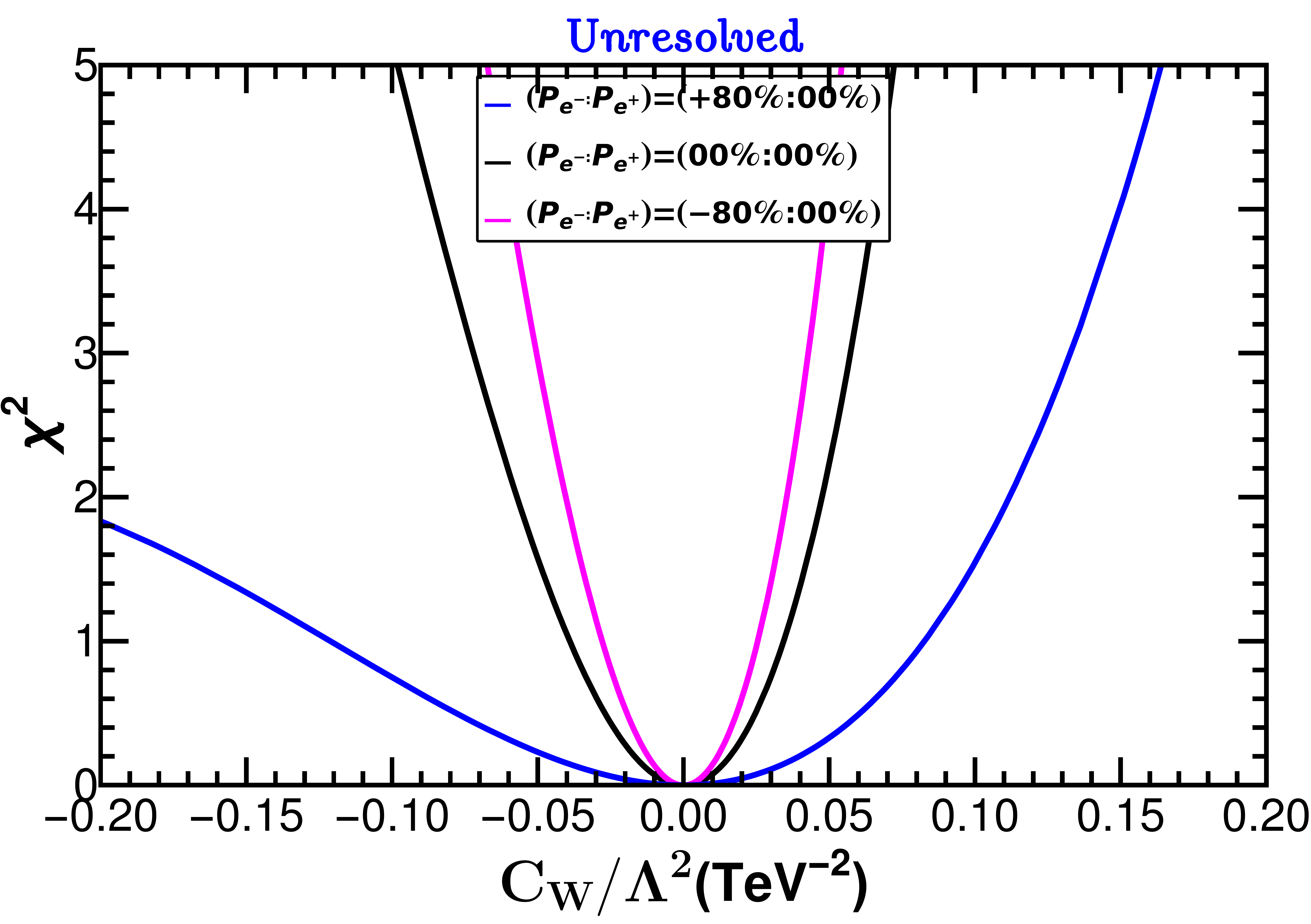}~
	\includegraphics[height=5cm,width=4.95cm]{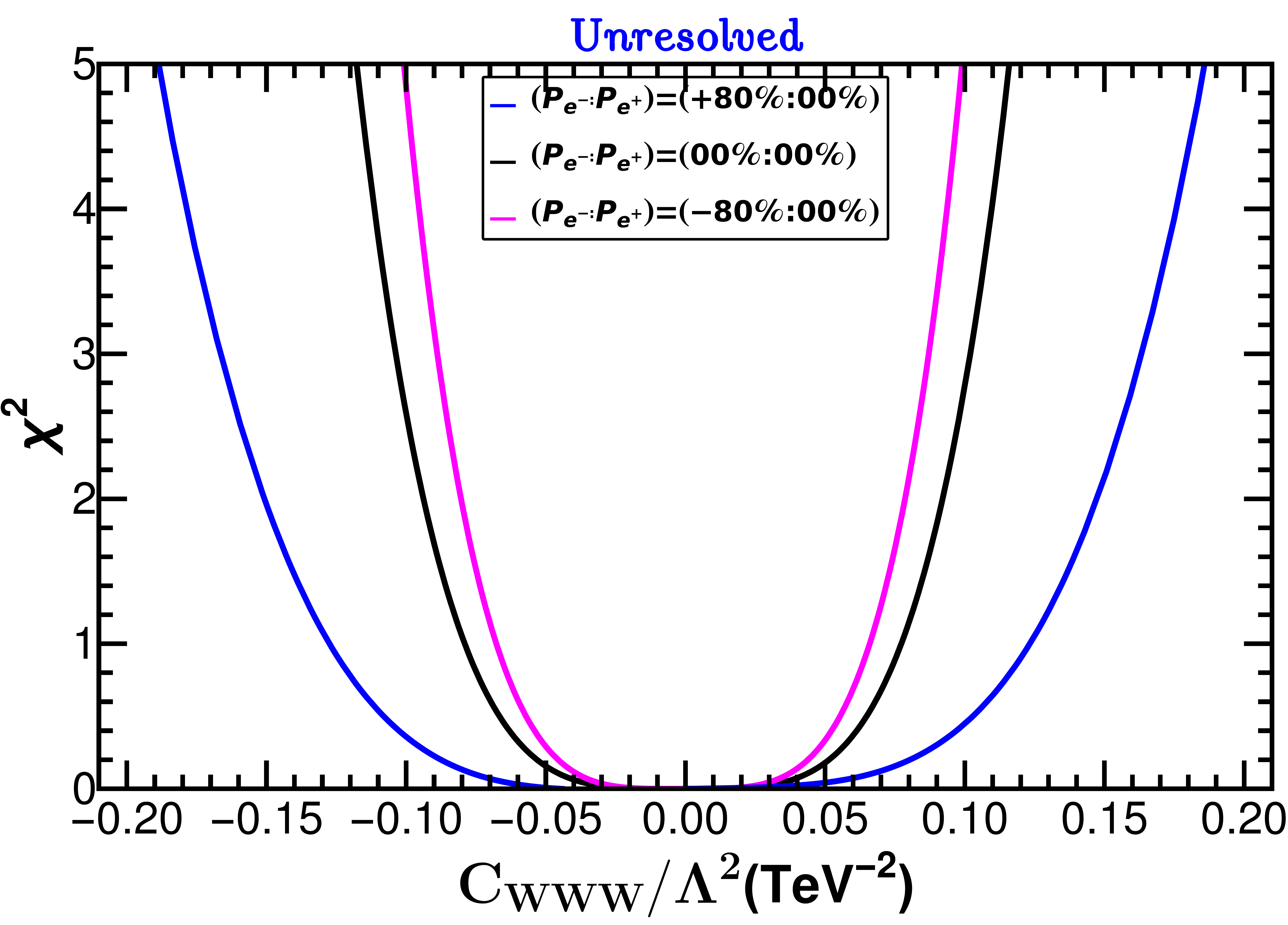}
	$$
	$$
	\includegraphics[height=5cm,width=4.95cm]{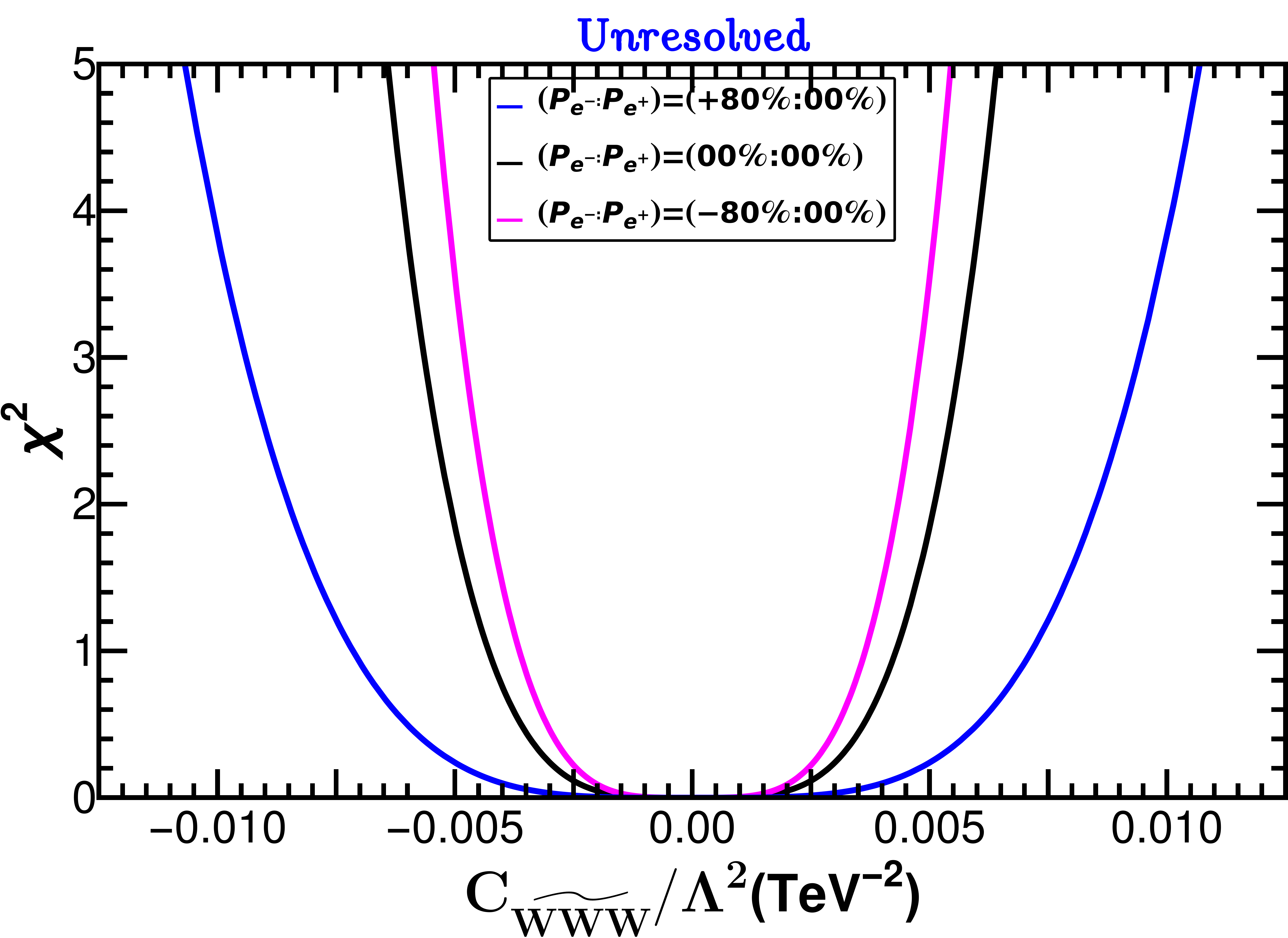}~~
	\includegraphics[height=5cm,width=4.95cm]{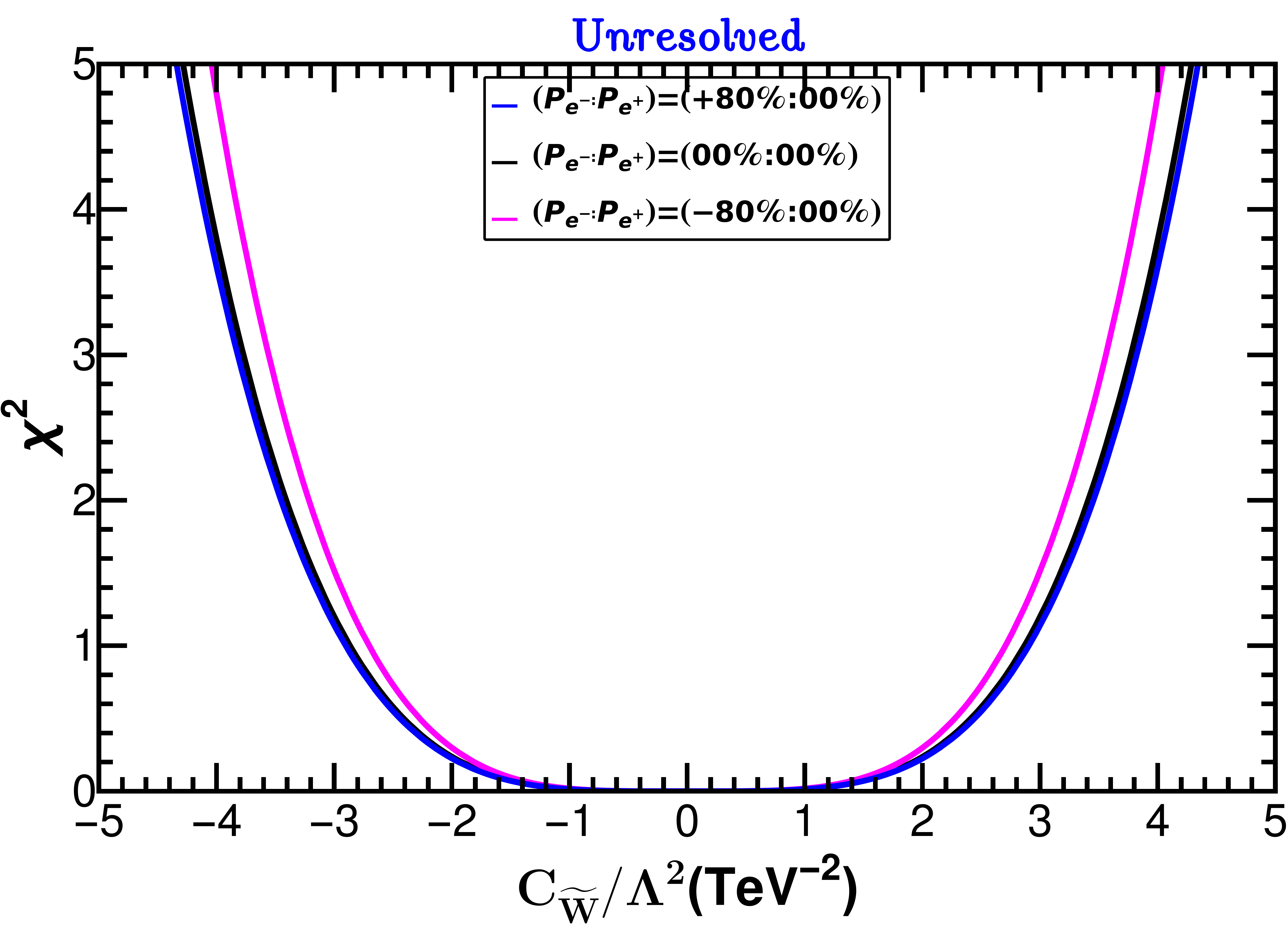}	
	$$
	\caption{Optimal $\chi^2$ variation with anomalous dimension-6 cTGCs for unresolved helicity combination of $W$ pair. Polarization information is written inside the plots.}
	\label{fig:chi2.pol.sum}
\end{figure}
%
\subsection{Optimal Sensitivity}
\label{sec:oot.ses}
\begin{figure}[t]
	$$   
	\includegraphics[height=5cm,width=4.95cm]{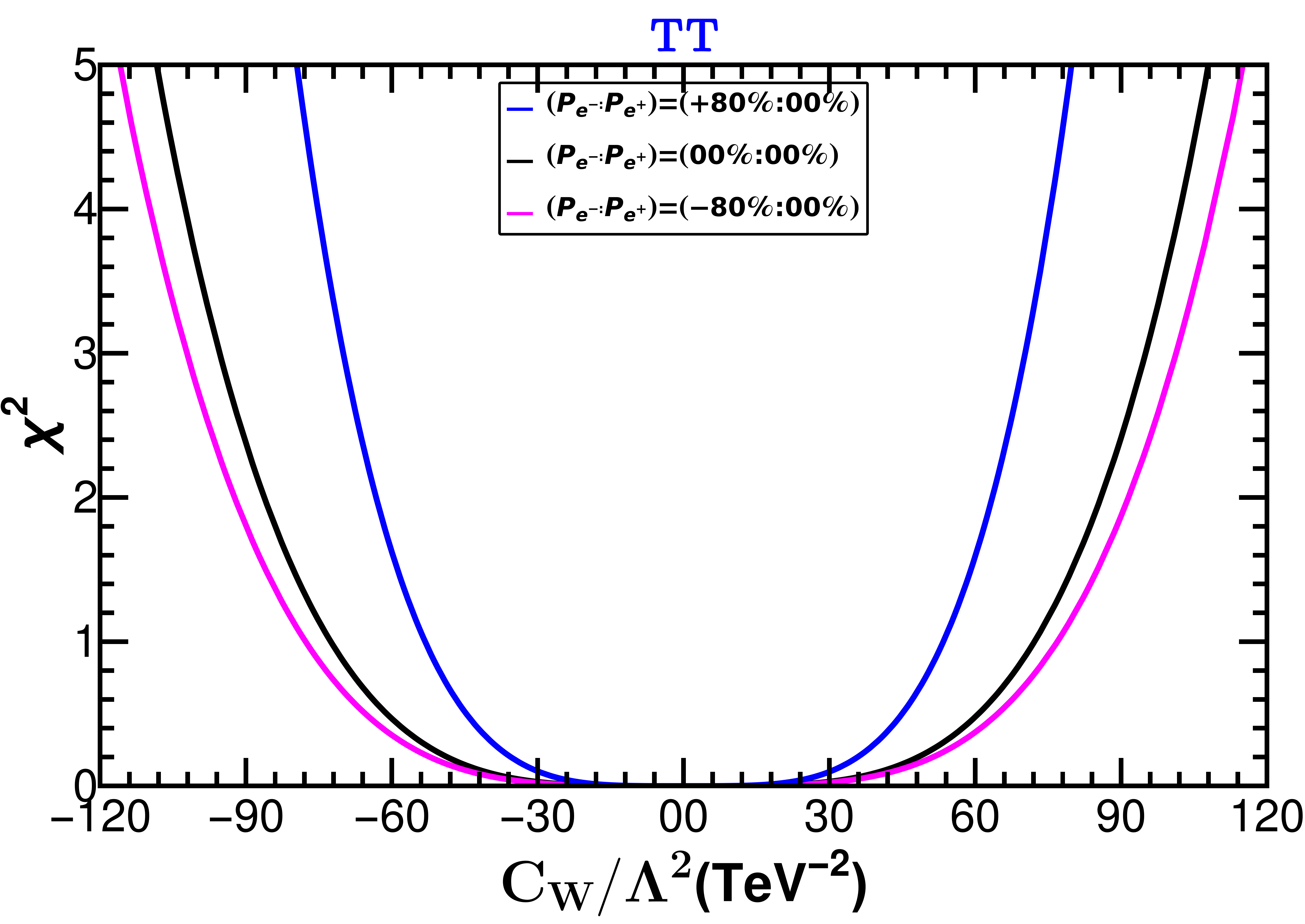}~~
	\includegraphics[height=5cm,width=4.95cm]{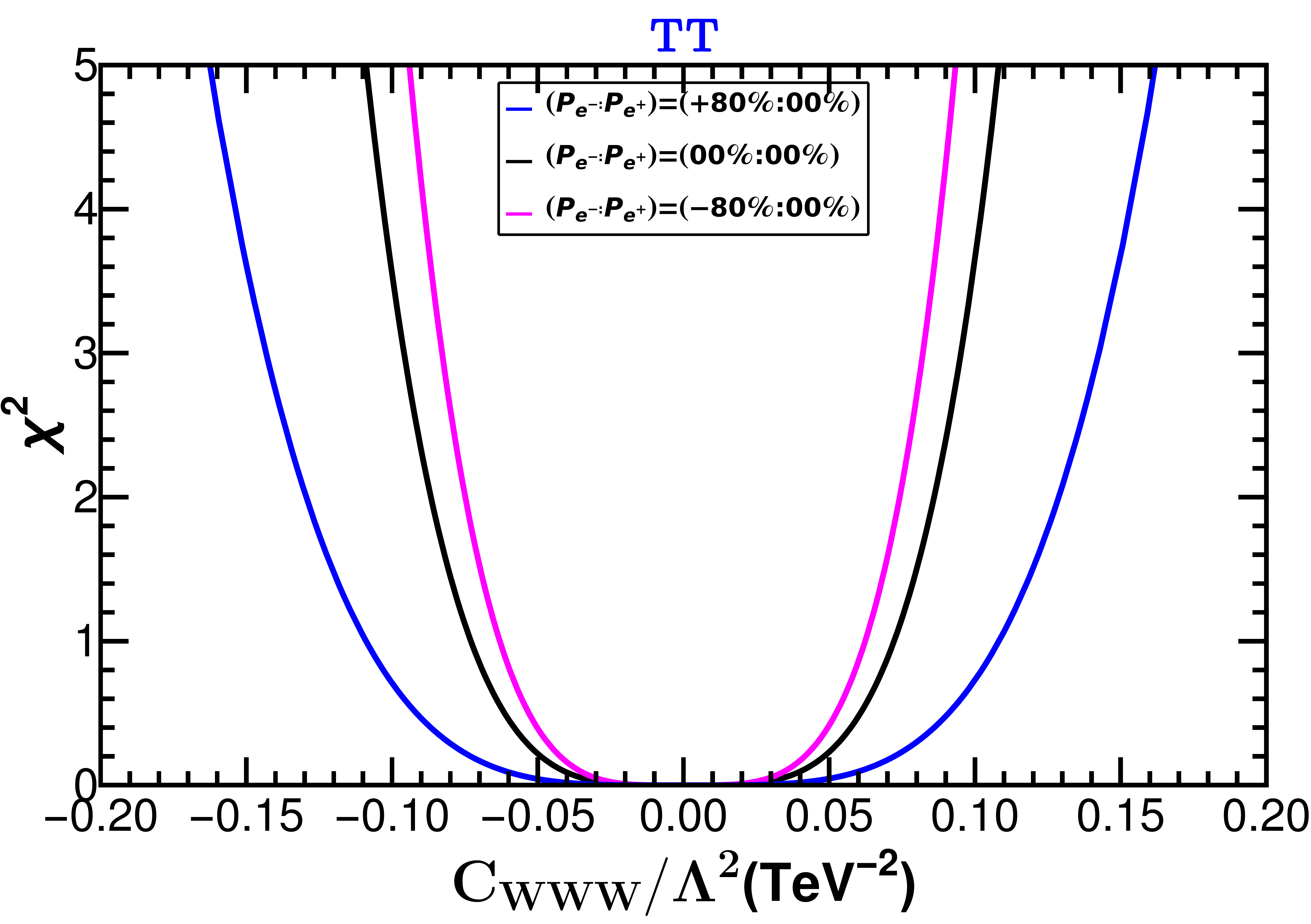}
	$$
	$$
	\includegraphics[height=5cm,width=4.95cm]{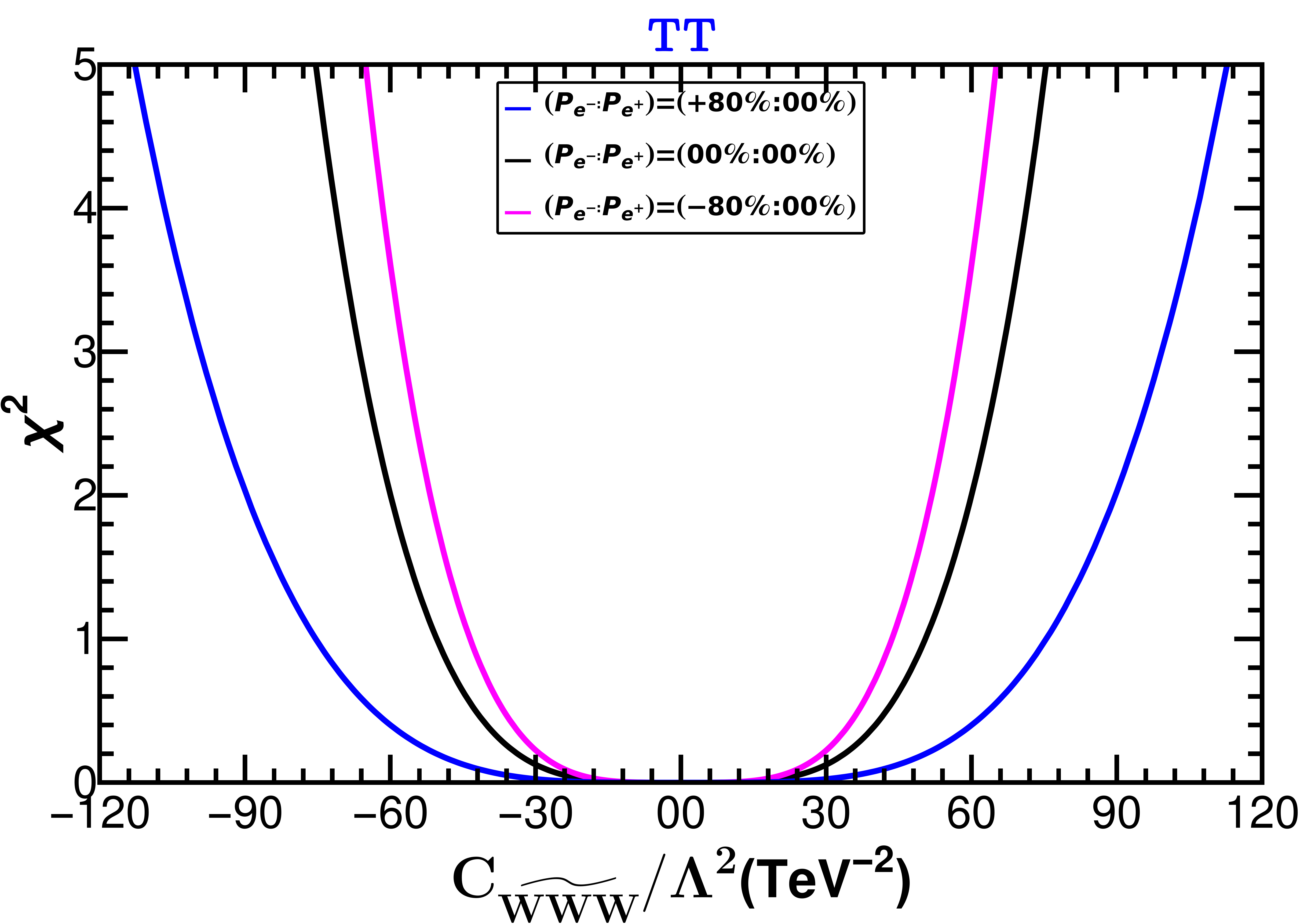}~~
	\includegraphics[height=5cm,width=4.95cm]{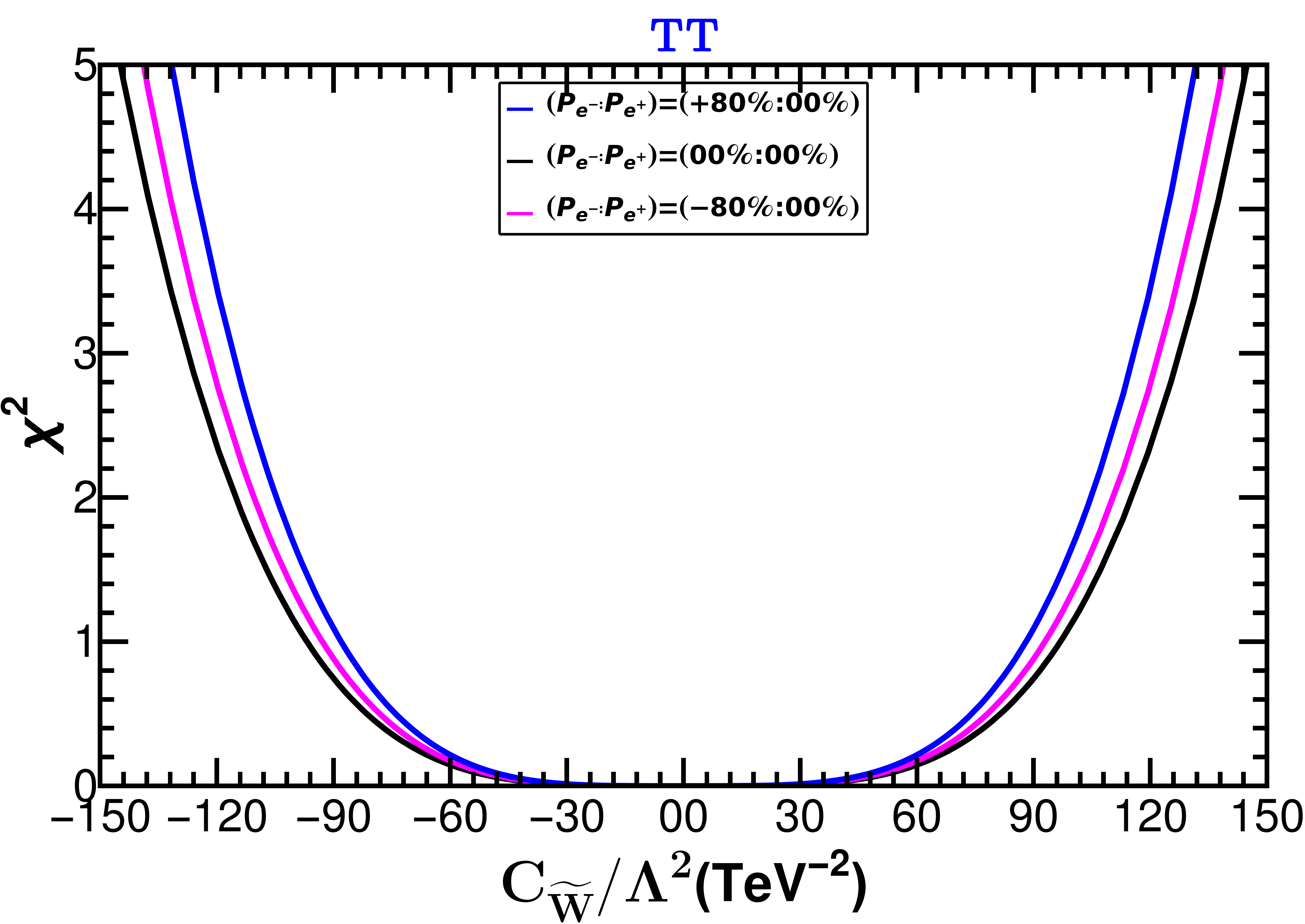}	
	$$
	\caption{Same as Fig.~\ref{fig:chi2.pol.sum} but for TT helicity combination of $W$ pair.}
	\label{fig:chi2.TT}
\end{figure}

\begin{figure}[H]
	$$
	\includegraphics[height=5cm,width=4.95cm]{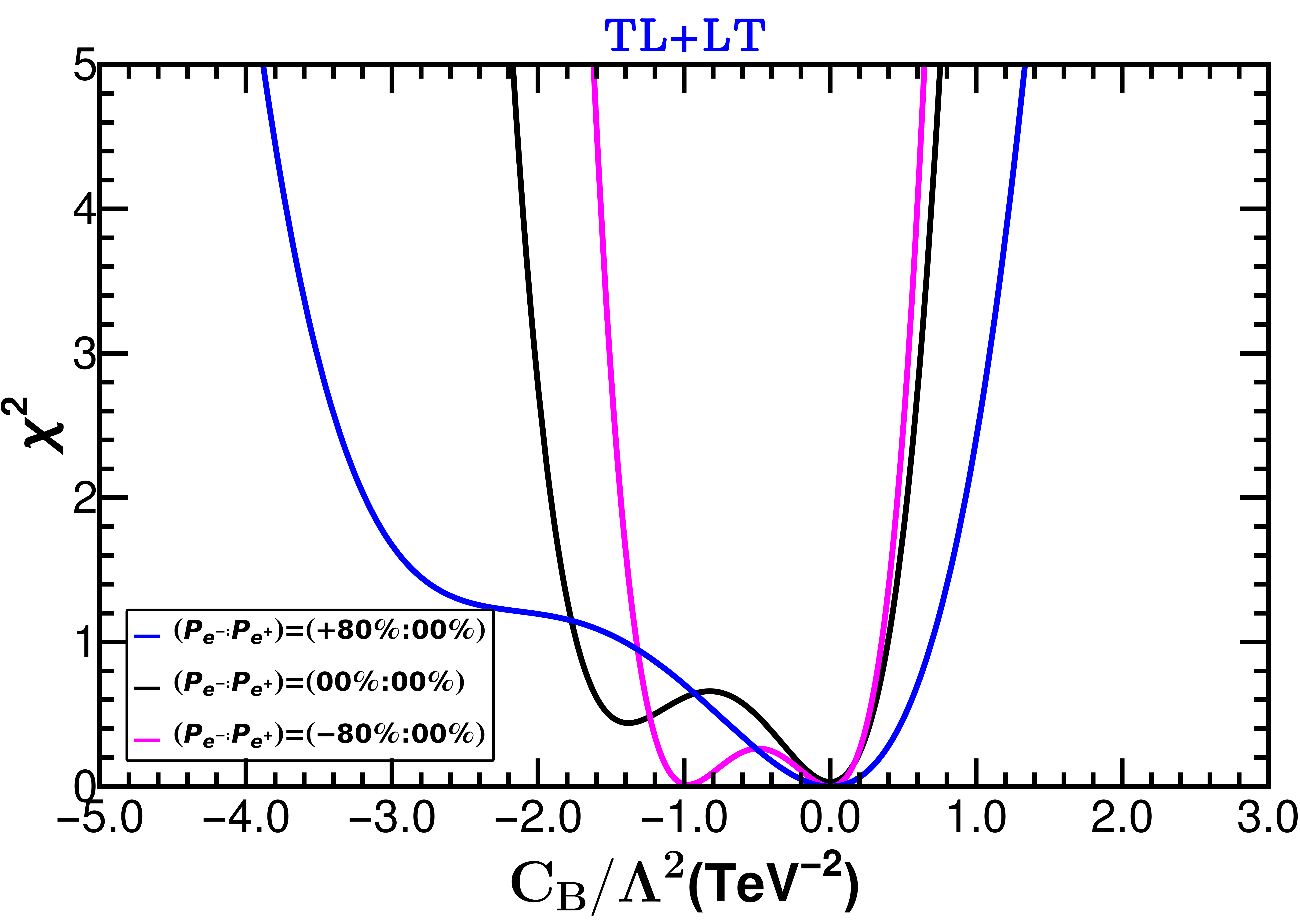}~
	\includegraphics[height=5cm,width=4.95cm]{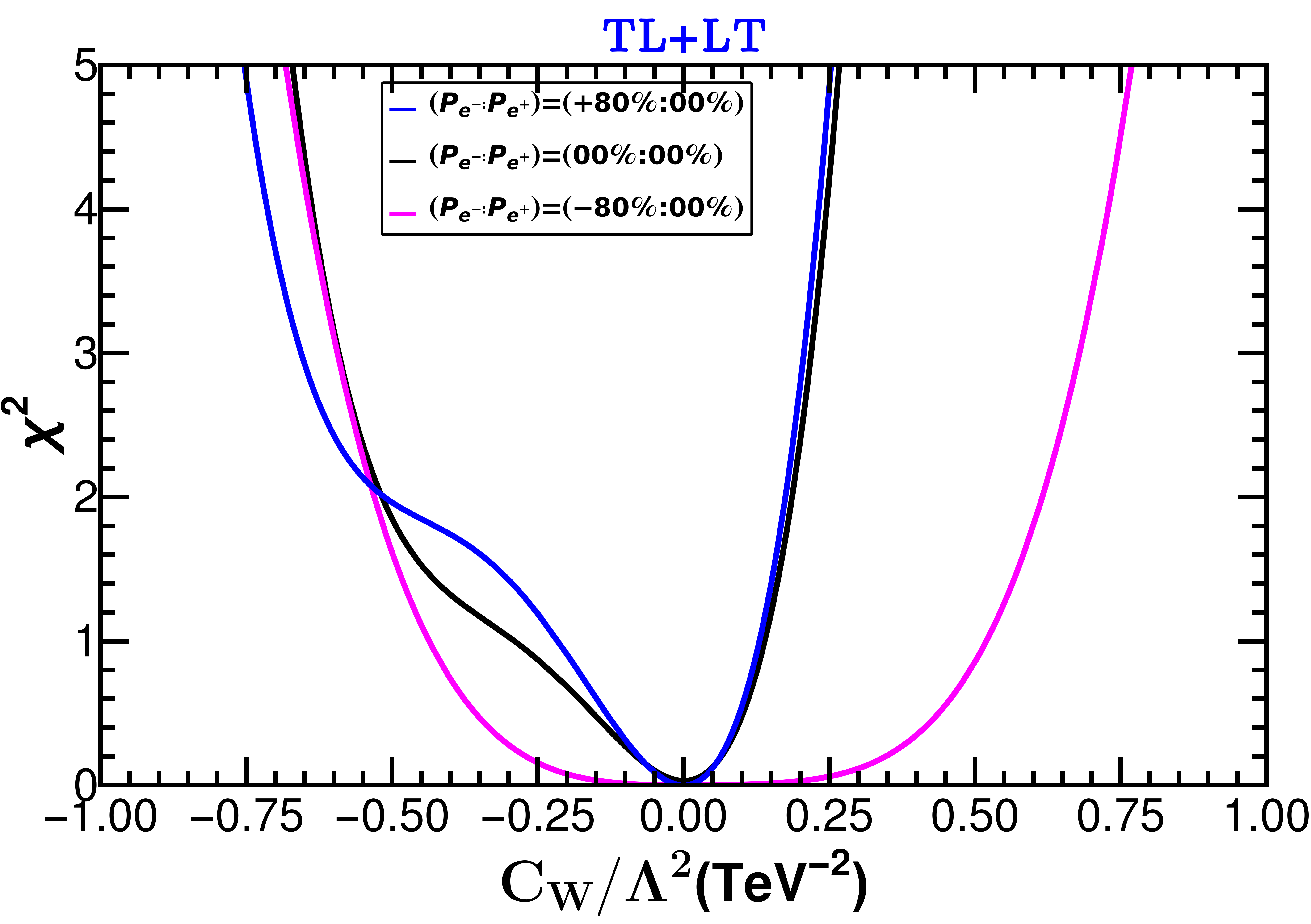}~
	\includegraphics[height=5cm,width=4.95cm]{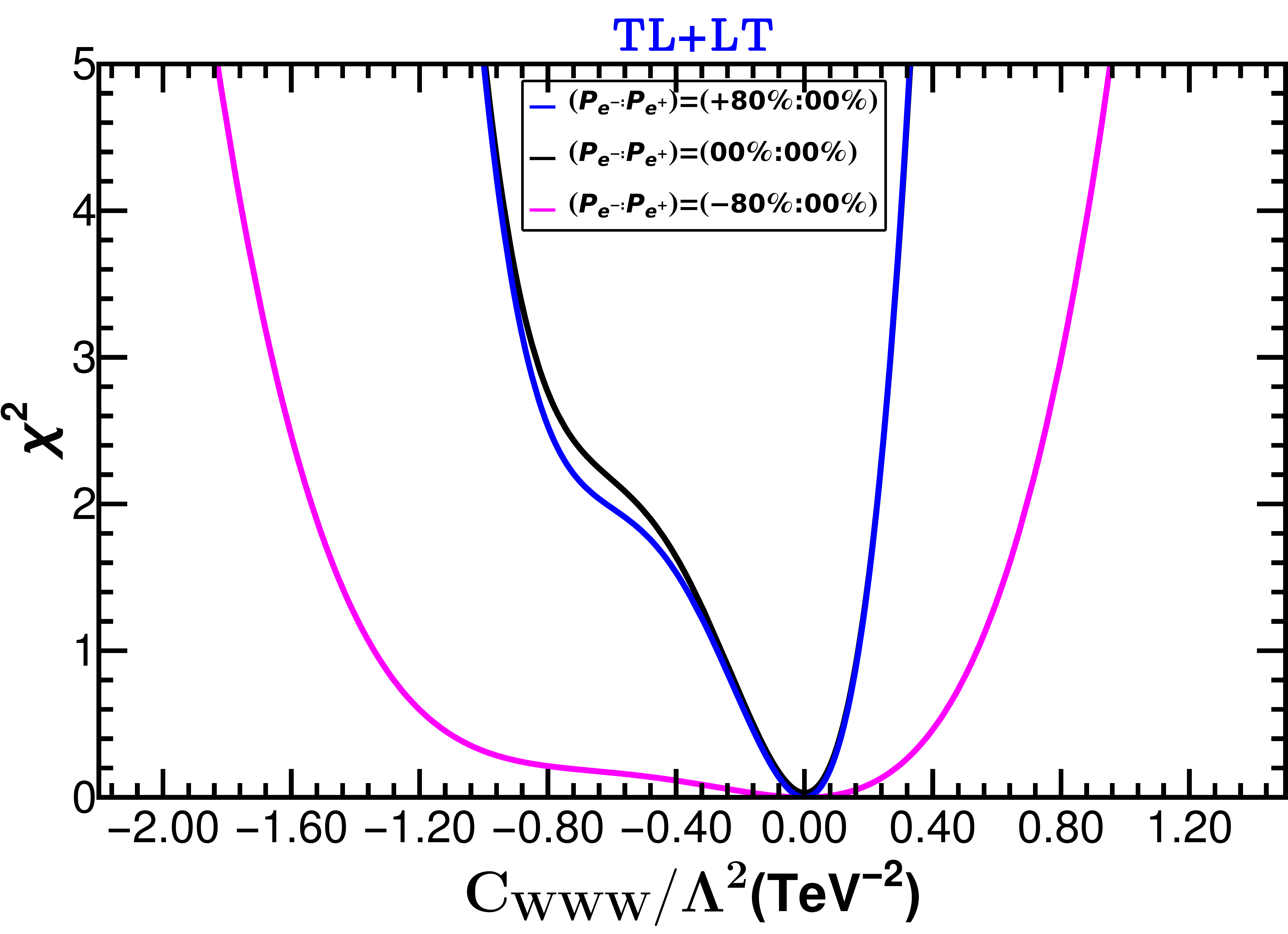}
	$$
	$$
	\includegraphics[height=5cm,width=4.95cm]{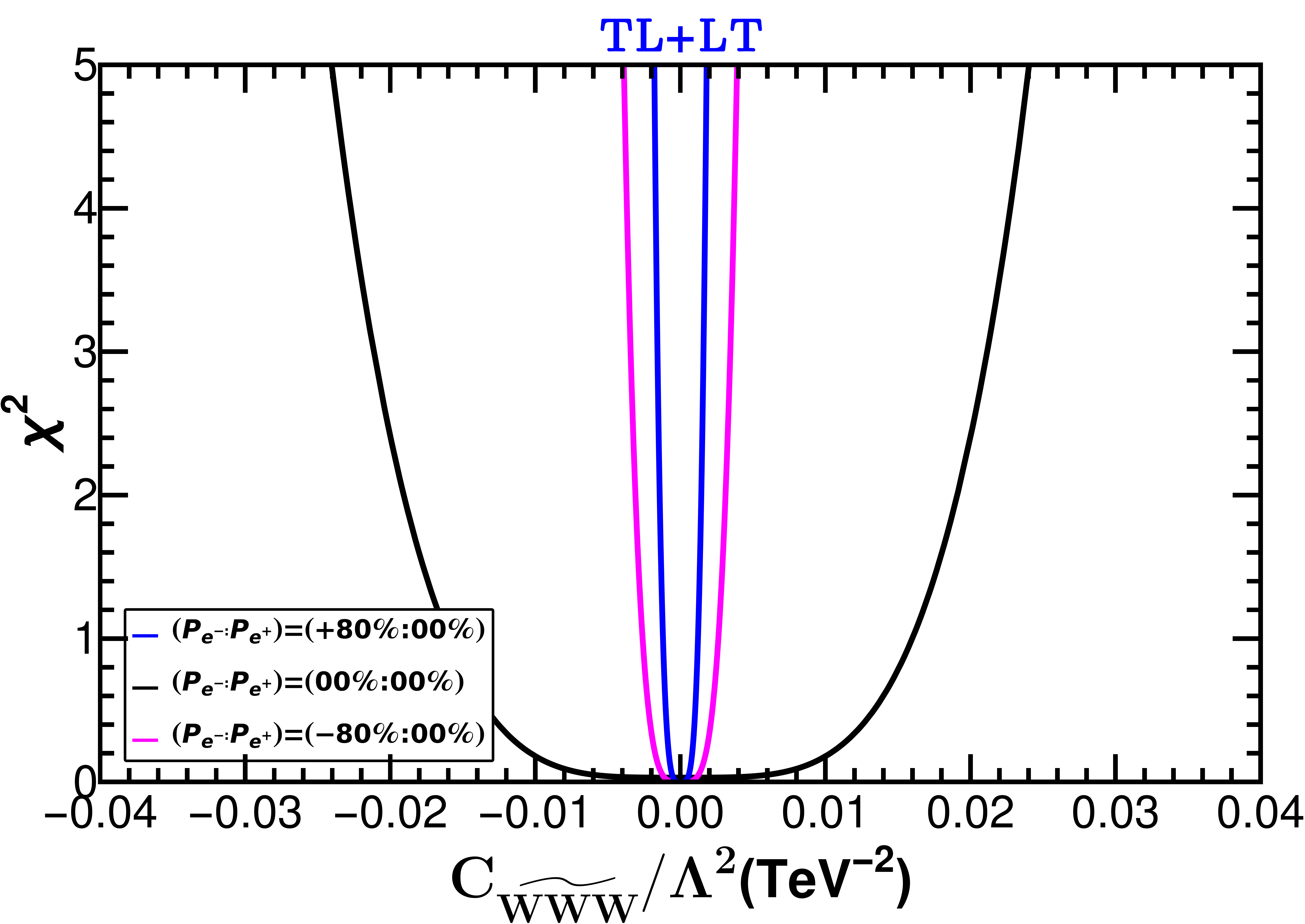}~
	\includegraphics[height=5cm,width=4.95cm]{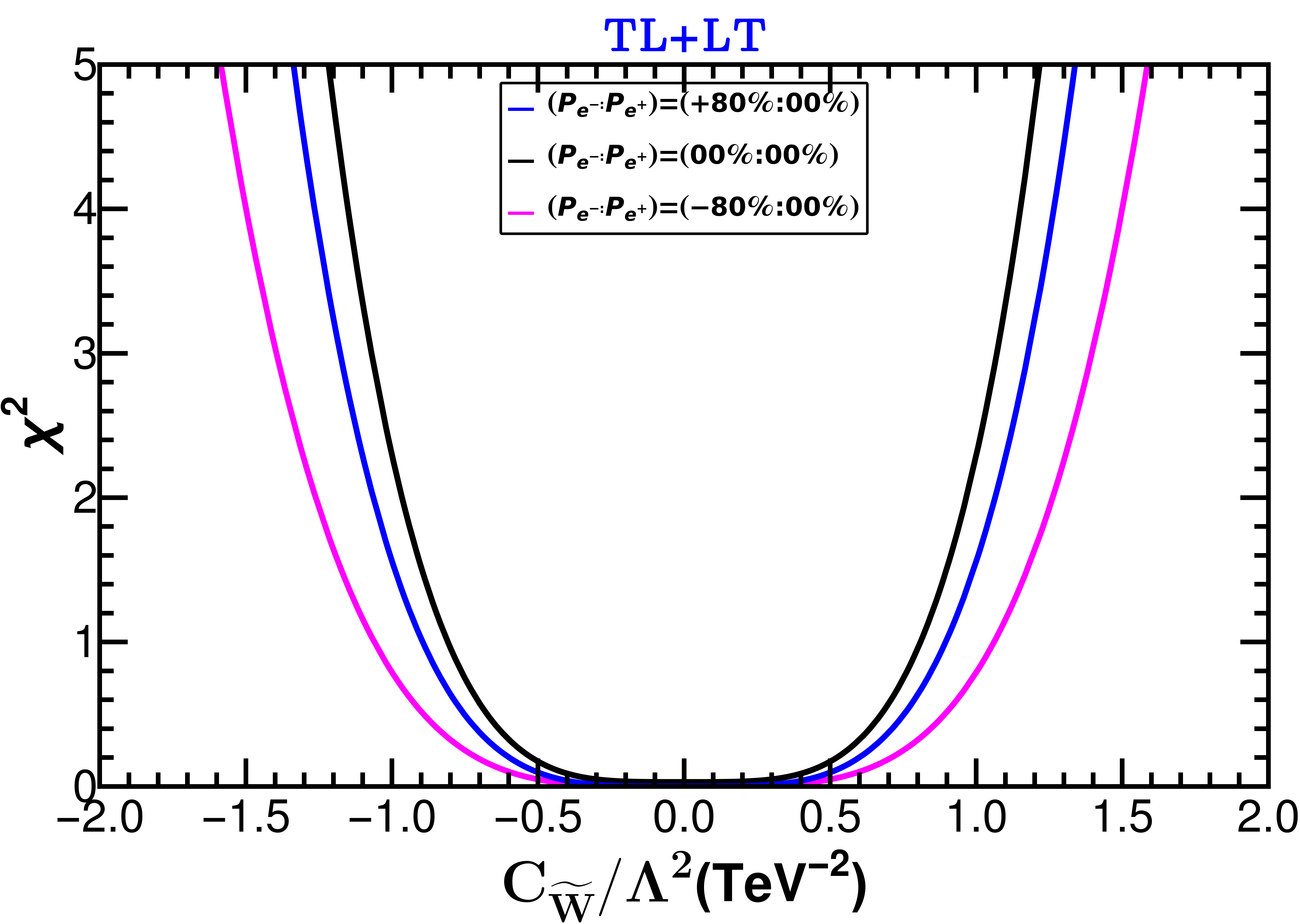}	
	$$
	\caption{Same as Fig.~\ref{fig:chi2.pol.sum} but for TL helicity combination of $W$ pair.} 	
	\label{fig:chi2.TL}
\end{figure}

\begin{figure}[t]
	\centering
	\includegraphics[height=5cm,width=4.95cm]{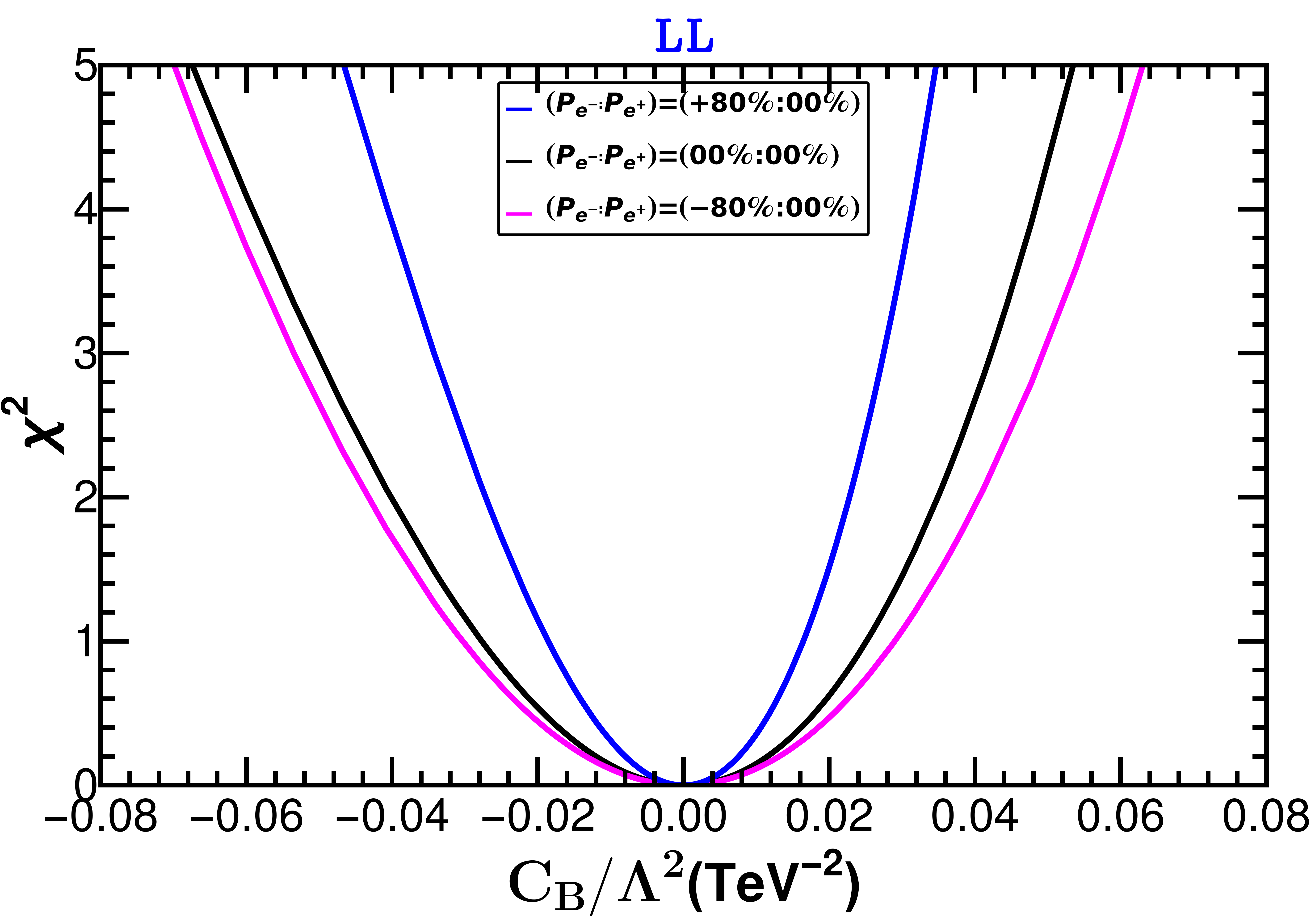} 
	\includegraphics[height=5cm,width=4.95cm]{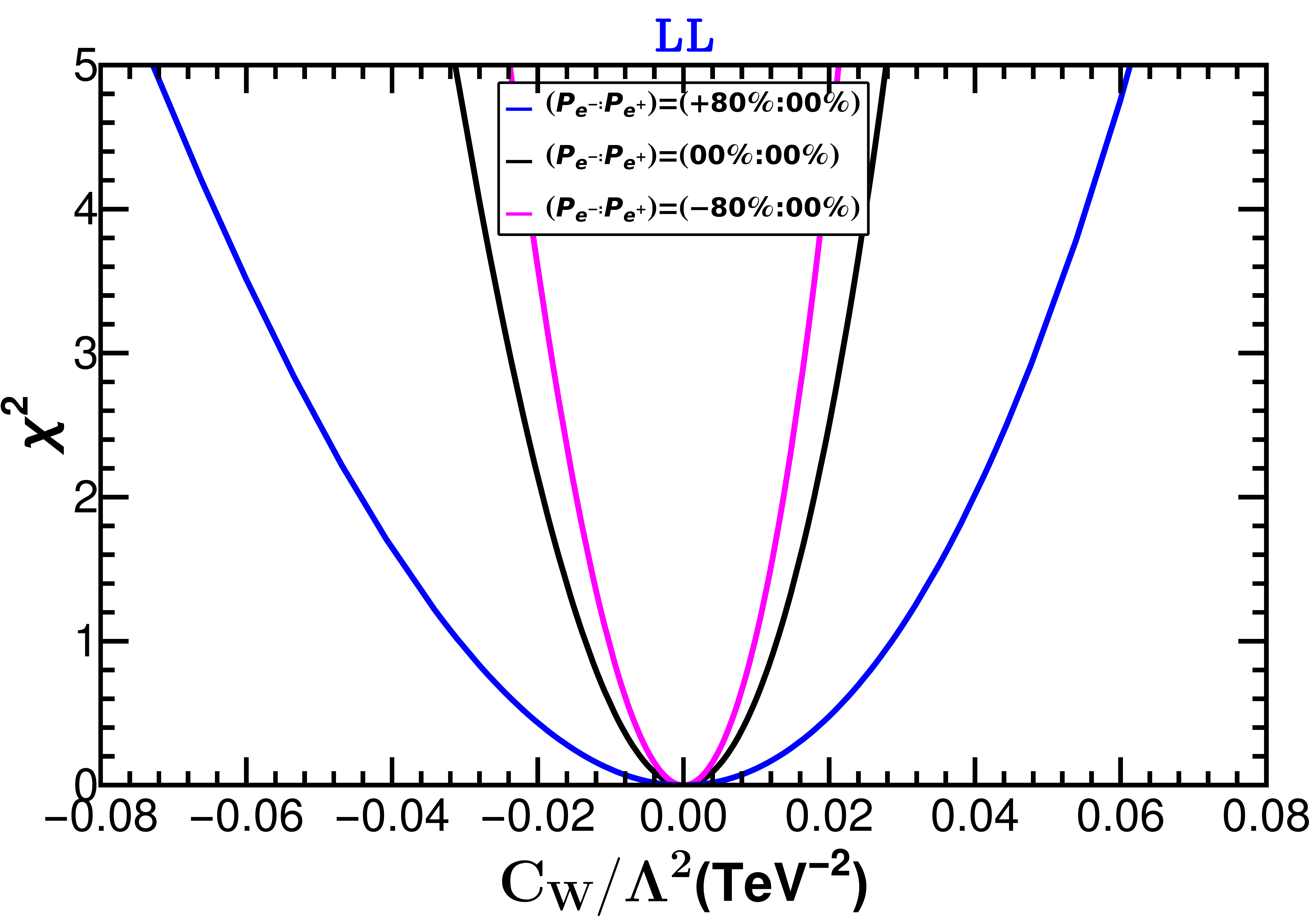} 		
	\includegraphics[height=5cm,width=4.95cm]{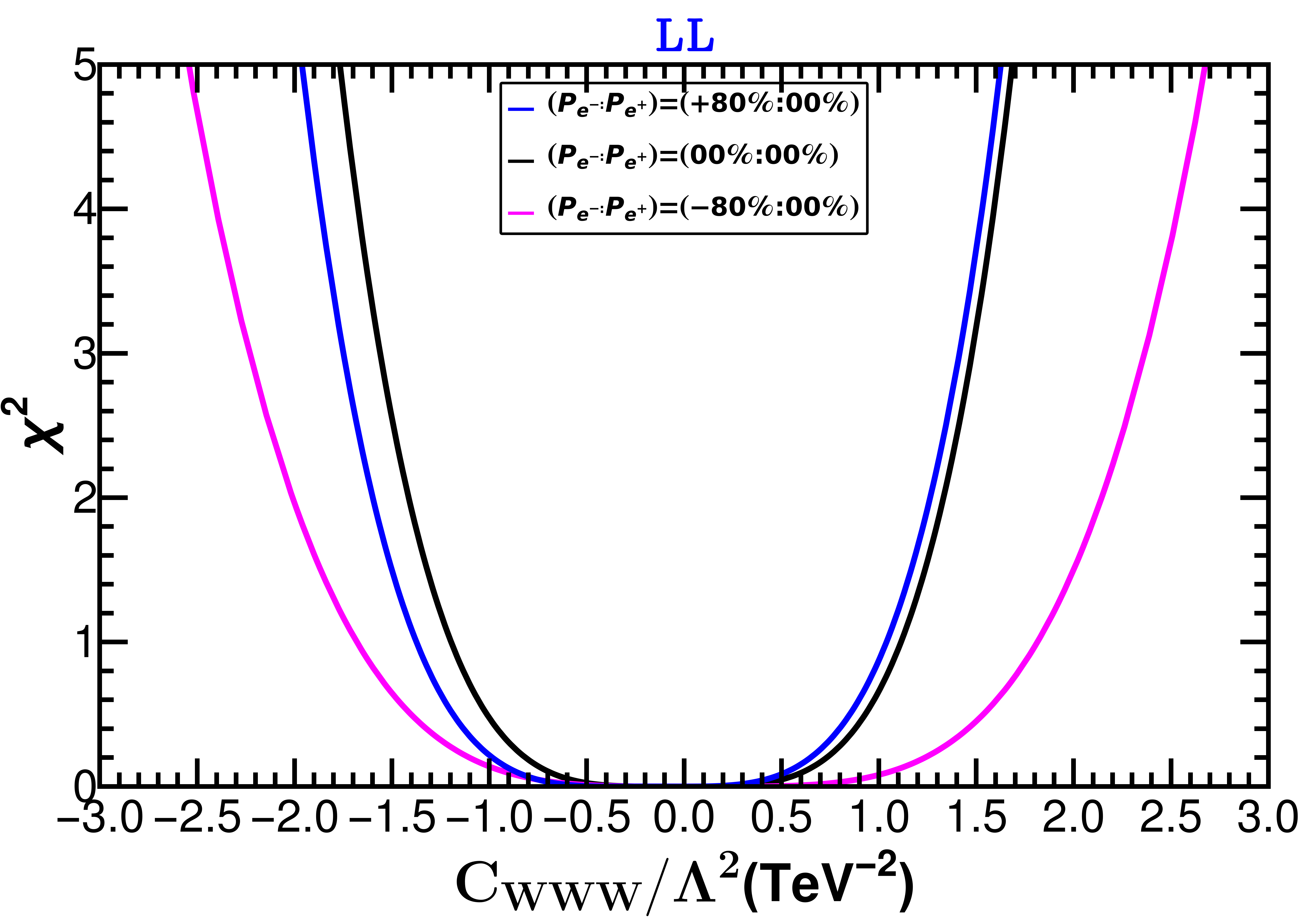} 
	\caption{Same as Fig.~\ref{fig:chi2.pol.sum} but for LL helicity combination of $W$ pair.} 	
	\label{fig:chi2.LL}
\end{figure}

\noindent
In this section, we explore the optimal sensitivity of different dimension-6 effective couplings contributing to the anomalous cTGCs with $\sqrt{s}=$ 3 TeV and $\mathfrak{L}_{\text{int}}=$ 1000 $\rm{fb}^{-1}$. Thanks to the distinct contribution to the cross-section, different helicity states of $W$ boson pair provide different sensitivity to the NP couplings. Judicious choice of initial beam polarization is also important to enhance the sensitivity of NP couplings by a few factors. Using Eq.~\eqref{eq:oot.chi.sqr}, the optimal $\chi^2$ variation with NP couplings for different final state helicity combinations considering three different polarization combinations are shown in Figs.~\ref{fig:chi2.pol.sum}-\ref{fig:chi2.LL}. The optimal sensitivity to NP couplings is influenced by the combined effect of two factors: the relative change in the contributions of the SM and NP to $WW$ production and the relative variation in the cross-section due to beam polarization.
For the unresolved helicity combination in Fig.~\ref{fig:chi2.pol.sum}, the optimal sensitivity for $C_W/\Lambda^2$ is highest with the polarization combination $\{P_{e^-}:P_{e^+}=-80\%:00\%\}$. This is because, under this polarization choice, the NP contribution to $WW$ production is enhanced. Similarly, the strongest constraint on $C_{WWW}/\Lambda^2$ is observed for the `TT' helicity combination (Fig.~\ref{fig:chi2.TT}), while the most stringent constraint on $C_{\widetilde{WWW}}/\Lambda^2$ occurs for the `LT' helicity combination (Fig.~\ref{fig:chi2.TL}). On the other hand, the opposite-sign polarization combination offers the best sensitivity for $C_B/\Lambda^2$ in the case of the `LL' helicity combination (Fig.~\ref{fig:chi2.LL}). In this scenario, SM contribution to the $WW$ cross-section is reduced due to the opposite-sign polarization choice. It is notable that, in the `TT' combination, apart from the $C_{WWW}/\Lambda^2$ case, the contributions from other couplings to $WW$ production at the same order are negligible. As a result, the constraints on these couplings are less stringent. Most stringent optimal limit (at 95\% CL) for each operator, corresponding helicity of $W$ boson pair, and initial beam polarization are summarized in Table~\ref{tab:oot.limit}. In comparison with the experimental limit listed in Table~\ref{tab:95cl.limit}, we find that the optimal sensitivities of $C_B/\Lambda^2$, $C_W/\Lambda^2$, and $C_{\widetilde{WWW}}/\Lambda^2$ are improved by two orders of magnitude. Meanwhile, for $C_{WWW}/\Lambda^2$ and $C_{\widetilde{W}}/\Lambda^2$,  an enhancement of one order of magnitude is achieved. 

\begin{table}[H]
	\centering
	{\renewcommand{\arraystretch}{1.3}%
		\begin{tabular}{ |c|c|c|c|c| } 
			\hline
			NP couplings & Optimal limits ({$\rm{TeV^{-2}}$}) & $WW$ helicity & beam polarization \\ 
			\hline
			$C_{B}/\Lambda^2$ & [$-0.041,+0.031$] & LL  & $\{P_{e^-}:P_{e^+}=+80\%:0\%\}$\\
			$C_{W}/\Lambda^2$ & [$-0.021,+0.019$] & LL & $\{P_{e^-}:P_{e^+}=-80\%:0\%\}$\\ 
			$C_{WWW}/\Lambda^2$ & [$-0.089,+0.088$] & TT & $\{P_{e^-}:P_{e^+}=-80\%:0\%\}$\\ 
			$C_{\widetilde{WWW}}/\Lambda^2$ & [$-0.002,+0.002$] & LT &$\{P_{e^-}:P_{e^+}=+80\%:0\%\}$\\ 
			$C_{\widetilde{W}}/\Lambda^2$ & [$-1.43,+1.43$] &LT & $\{P_{e^-}:P_{e^+}=+80\%:0\%\}$\\
			\hline
	\end{tabular}}
	\caption{Most stringent optimal sensitivity (95\% CL) fo dimension-6 cTGCs at the CLIC with $\sqrt{s}=3$ TeV and $\mathfrak{L}_{\text{int}}=1000~\text{fb}^{-1}$.}
	\label{tab:oot.limit}
\end{table}
\subsection{Sensitivity comparison: OOT vs standard $\chi^2$ analysis}
\label{sec:}
\begin{figure}[t]
	$$
	\includegraphics[height=5cm,width=4.95cm]{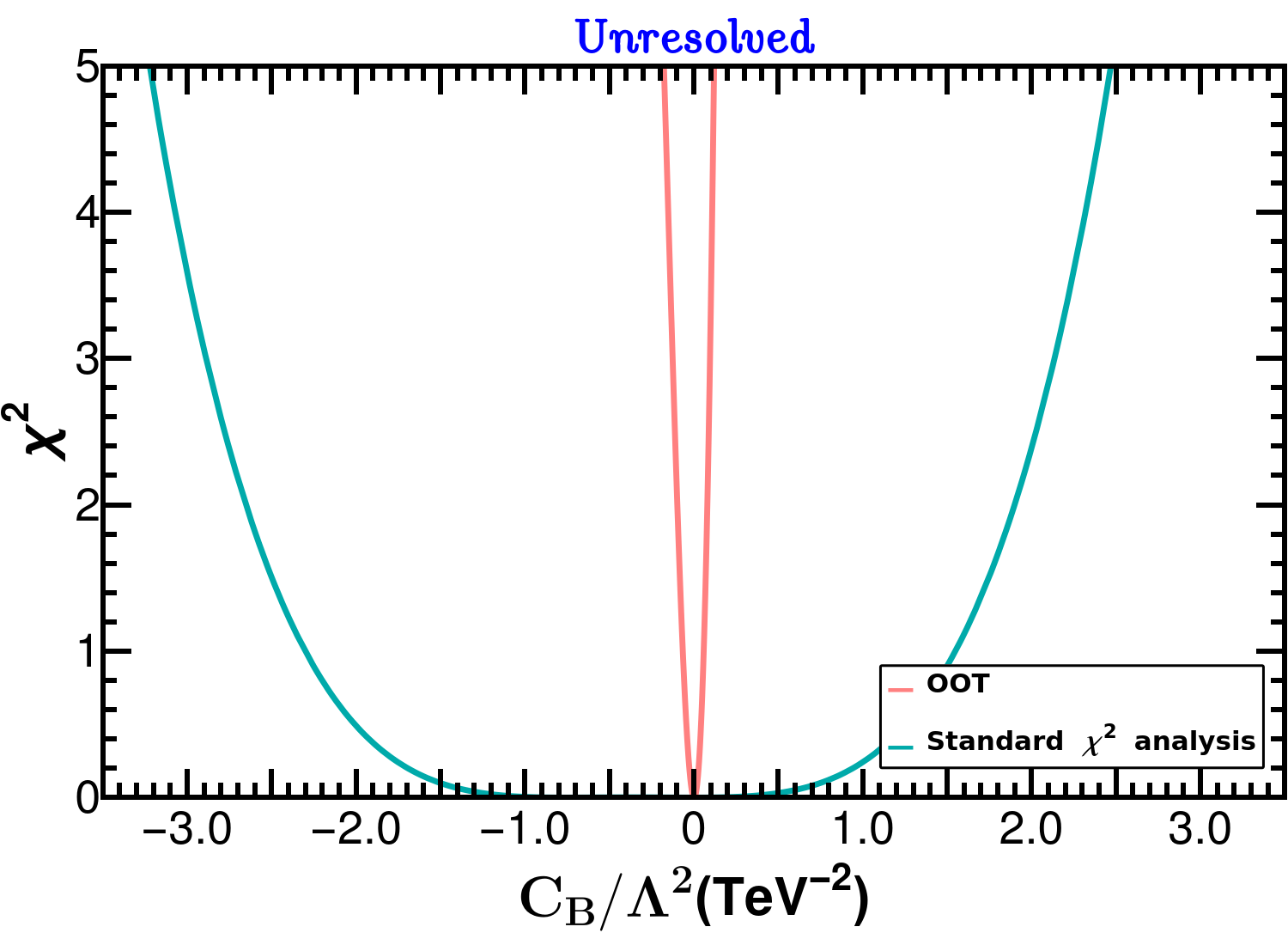}~
	\includegraphics[height=5cm,width=4.95cm]{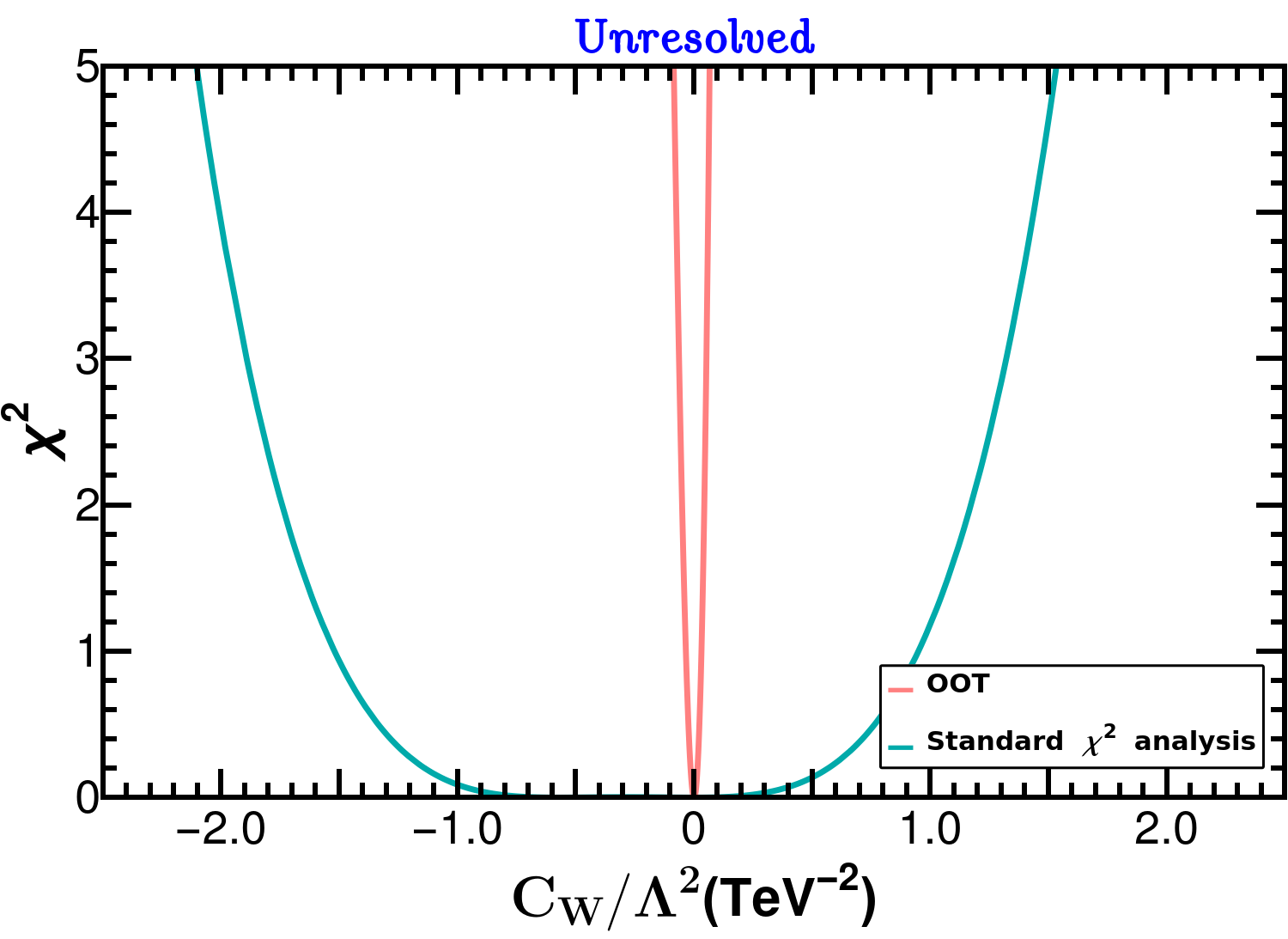}~	
	\includegraphics[height=5cm,width=4.95cm]{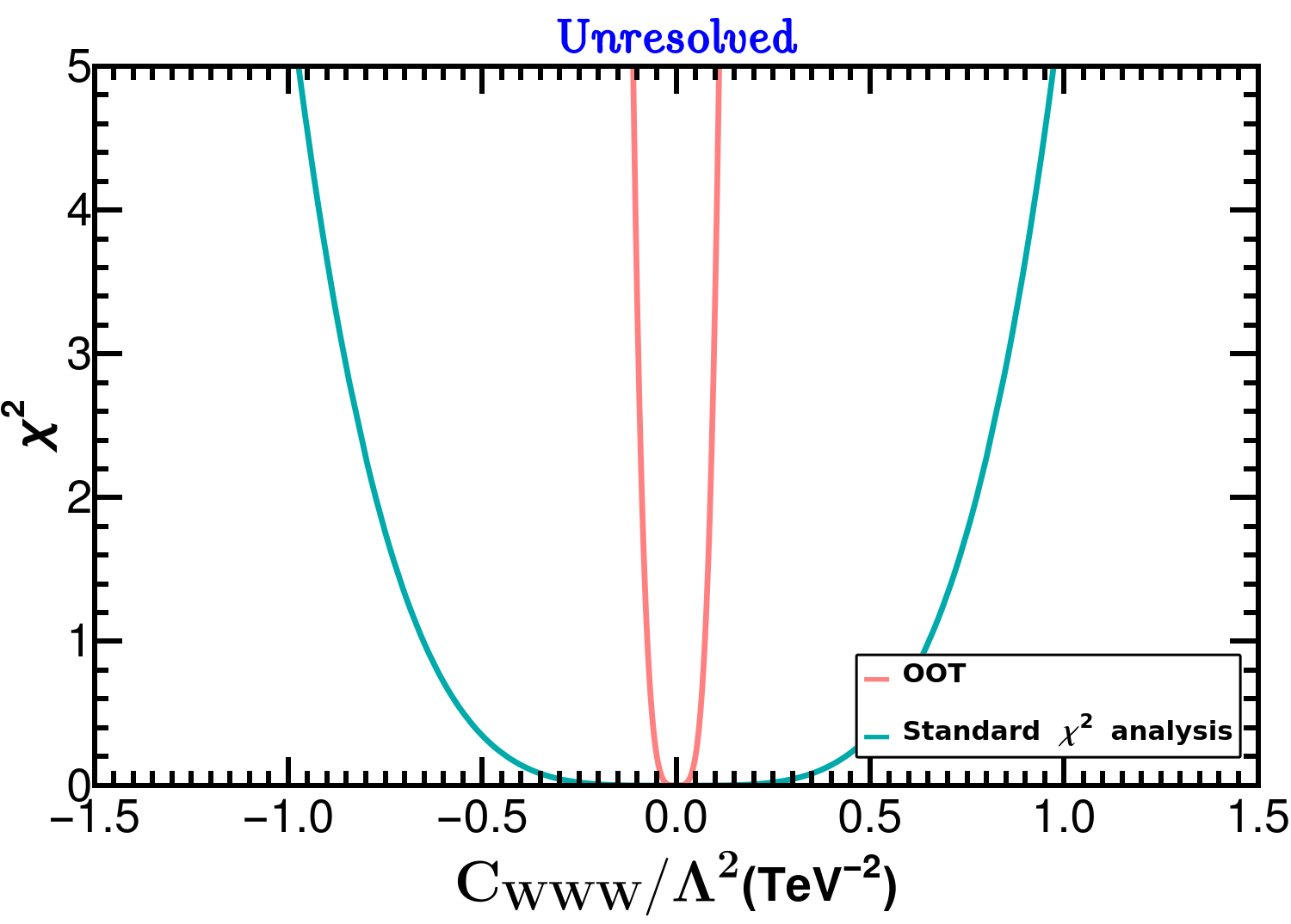}
	$$
	$$
	\includegraphics[height=5cm,width=4.95cm]{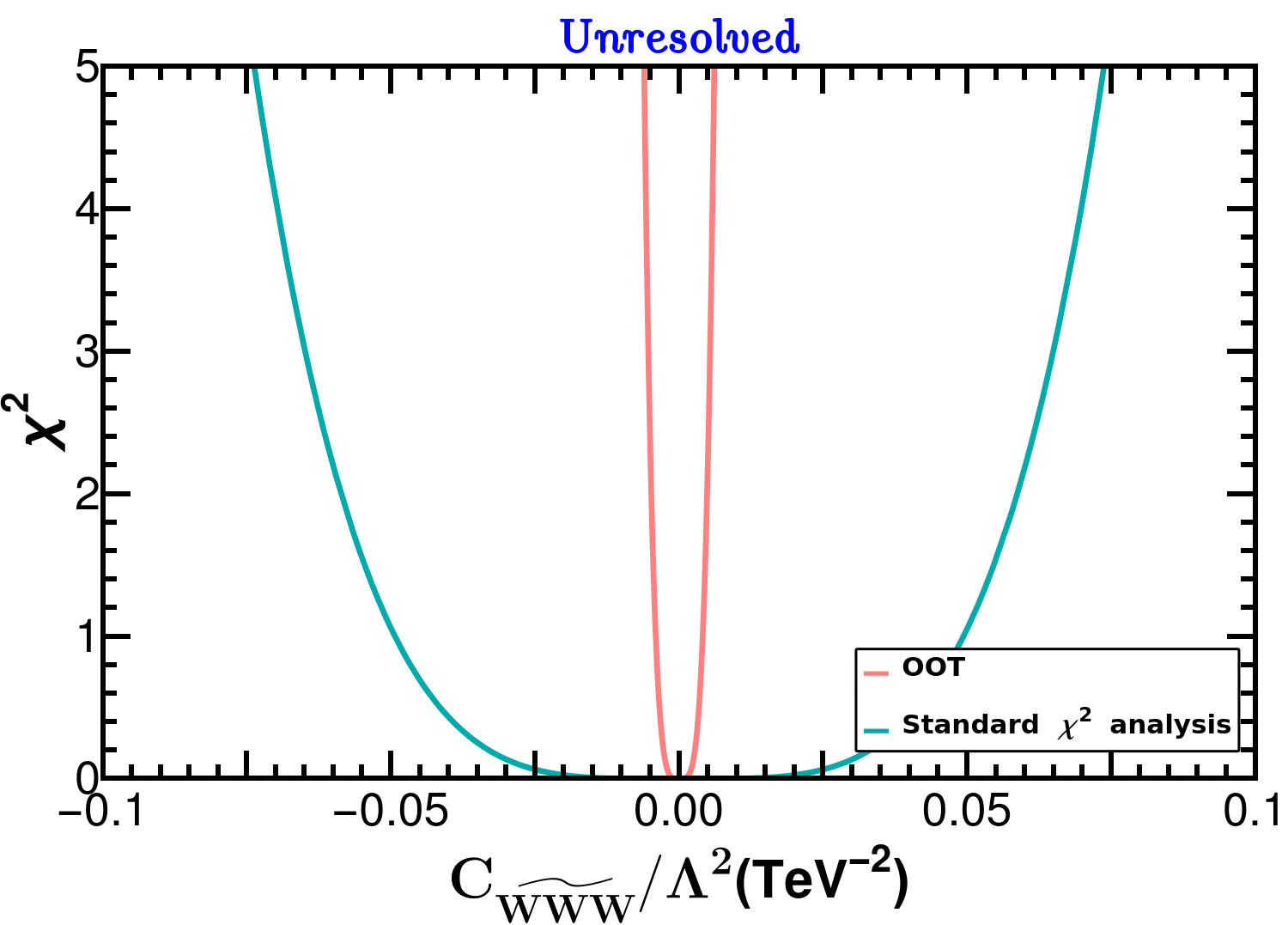}~~
	\includegraphics[height=5cm,width=4.95cm]{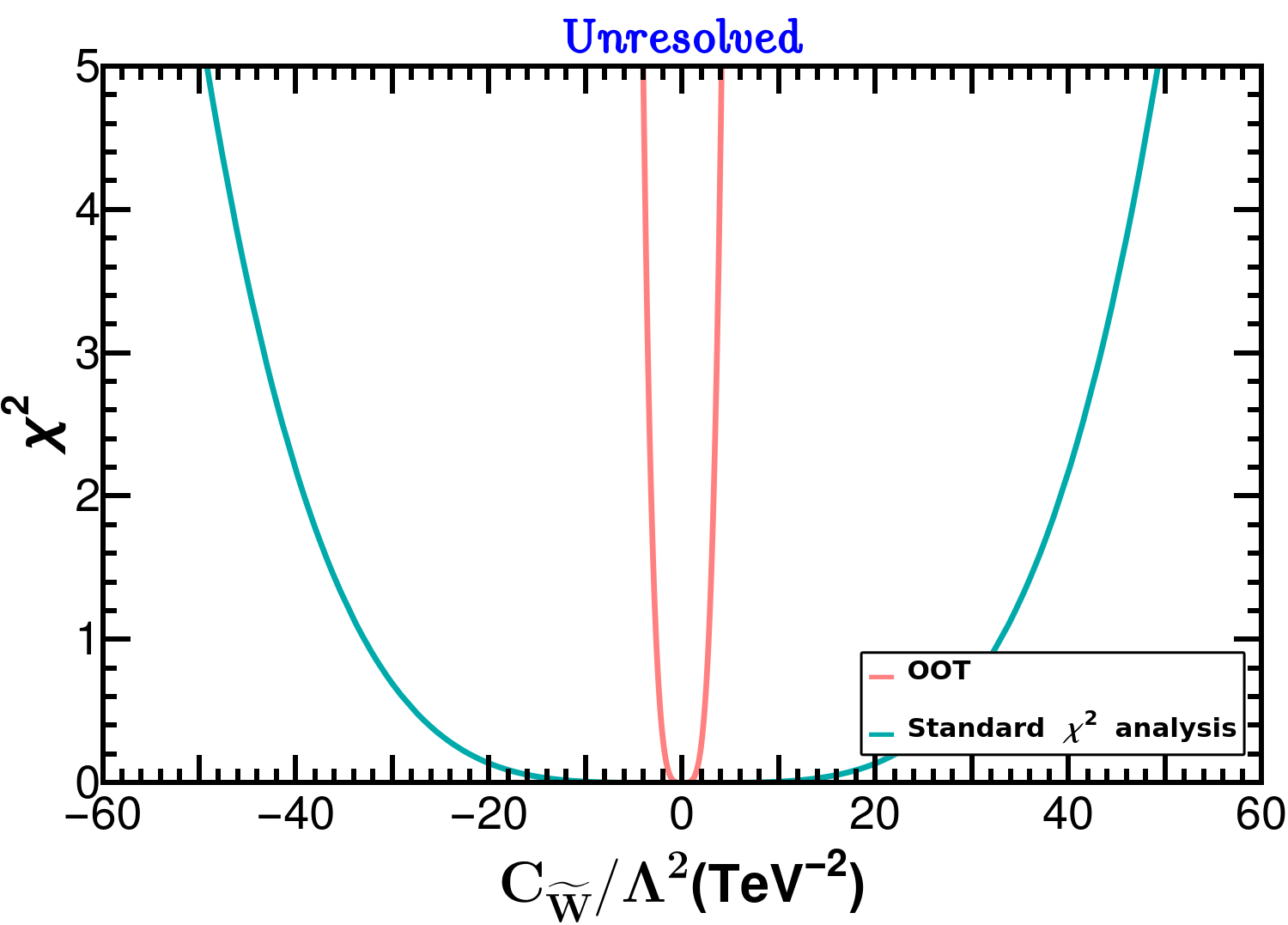}	
	$$
	\caption{Comparison of $\chi^2$ variation between OOT and standard scenario for different dimension-6 effective couplings. We consider the unresolved helicity state of $W$ and unpolarized initial beam.} 
	\label{fig:chi2.comp}
\end{figure}

\noindent
We turn to discuss the utility of the OOT over standard $\chi^2$ analysis in estimating the sensitivity of dim-6 effective couplings. To this end, the definition of the standard $\chi^2$ function is given by
\begin{equation}
	\chi^2 =\sum^{\rm{bins}}_{j} \left(\frac{N_j^{\tt obs}-N_j^{\tt theo}(g_i)}{\Delta N_j}\right)^2,
	\label{eq:chi2.trad}
\end{equation}
where $N_j^{\tt obs}$ and $N_j^{\tt theo}$ represent the number of observed/simulated events and theoretical prediction, respectively, in the $\rm{j^{\tt th}}$ bin of the differential cross-section distribution. The statistical uncertainty in the $\rm{j^{\tt th}}$ bin, denoted by $\Delta N_j$, is defined as $\Delta N_j = \sqrt{N^{\tt obs}_j}$. Using the specified CLIC design mentioned above, we show the standard $\chi^2$ variation using Eq.~\eqref{eq:chi2.trad}, shown in Fig.~\ref{fig:chi2.comp} in cyan color for the unresolved helicity state of $W$ boson pair with unpolarized beam. We also display the optimal $\chi^2$ variation in the same plot with pink color for comparison. A significant improvement is achieved in estimating the sensitivity of NP couplings in the case of the OOT. We find that the optimal sensitivity of $C_{B}/\Lambda^2$, $C_{W}/\Lambda^2$, and $C_{WWW}/\Lambda^2$ improves over the traditional sensitivity by factors of 20, 25, and 9, respectively, whereas for both CP-violating scenarios, optimal sensitivity is better for both cases by a factor of 12.

In addition to statistical uncertainties, systematic uncertainties can significantly impact the sensitivity to the dimension-6 NP couplings. To account for this, the total uncertainty in the denominator of Eq.~\eqref{eq:chi2.trad} is modified as $\Delta N_j \sqrt{(1 + r^2 N_j^{\tt obs})}$, where $r$ denotes the fractional systematic uncertainty relative to the statistical uncertainty. Including systematic uncertainties of 5\%, 10\%, and 20\% leads to a reduction in the sensitivity to the dimension-6 NP couplings by approximately 3\%, 8\%, and 24\%, respectively.

%
\section{Constraints from Electric Dipole Moment}
\label{sec:edm}
\noindent
Electric dipole moment (EDM) proves to be one of the cleanest probes of NP signal stemming from CP-violation. A non-zero EDM arises from CP violation, and the contribution from the SM is highly suppressed. Recent experiments have probed the EDM of electrons and neutrons with unprecedented precision. This makes the EDM an exceptionally sensitive probe for uncovering BSM physics. NP at the TeV scale, accompanied by CP-violating interactions, could produce sizable EDM that are within reach of near-future observations. The most recent ACME bound on electron EDM turns out to be \cite{ACME:2018yjb}:
\begin{equation}
	|d_e| < 1.1 \times 10^{-29}~\rm{e.cm}~~~or~~~ |d_e| < 1.7 \times 10^{-16}~GeV^{-1}.
	\label{eq:edm.cons}
\end{equation}
Since the EDM measurement has been performed at remarkable precision, they are competitive with collider measurements in constraining CP-violating higher dimensional effective operators. 

From Eq.~\eqref{eq:dim6.ops}, we identify two dimension-6 CP-violating SMEFT operators that contribute to cTCGs and also potentially induce electron EDM. Therefore, it is essential in this context to examine the constraints on these operators derived from the aforementioned EDM experiment. The contributions of ${\mathscr O}_{\widetilde{WWW}}$ and ${\mathscr O}_{\widetilde{W}}$ operators to electron EDM can be written as follows~\cite{Panico:2018hal}
\begin{equation}
	\frac{d_e}{e} = \frac{y_e g m_w}{32 \sqrt{2} \pi^2 \Lambda^2} \left(3 s_w g^2 C_{\widetilde{WWW}}+\ln{\left[\frac{m_h}{\Lambda_C}\right]} C_{\widetilde{W}}\right),
	\label{eq:edm}
\end{equation}

\begin{figure}[t]
	$$
	\includegraphics[height=5cm,width=4.95cm]{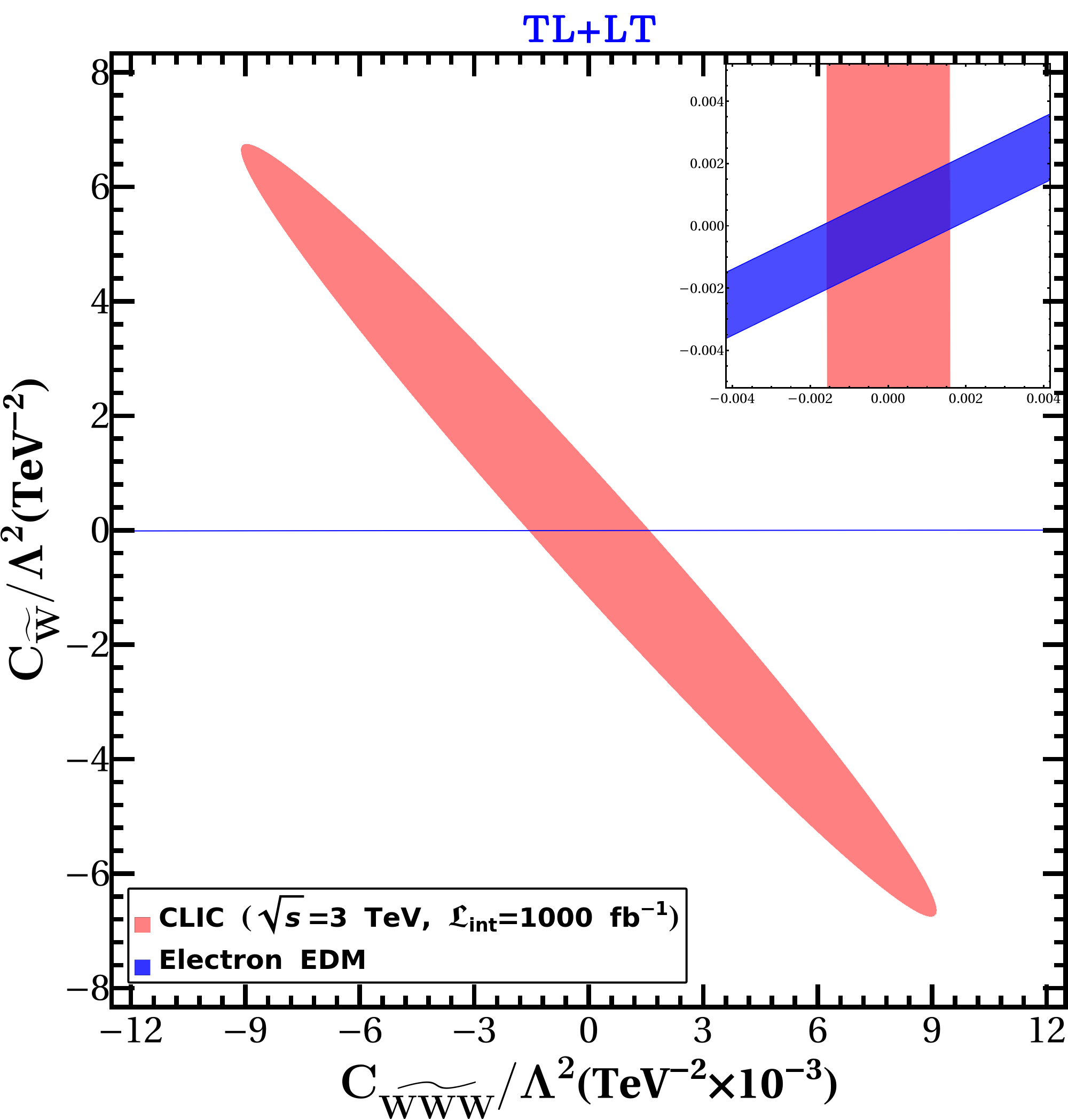}~
	\includegraphics[height=5cm,width=4.95cm]{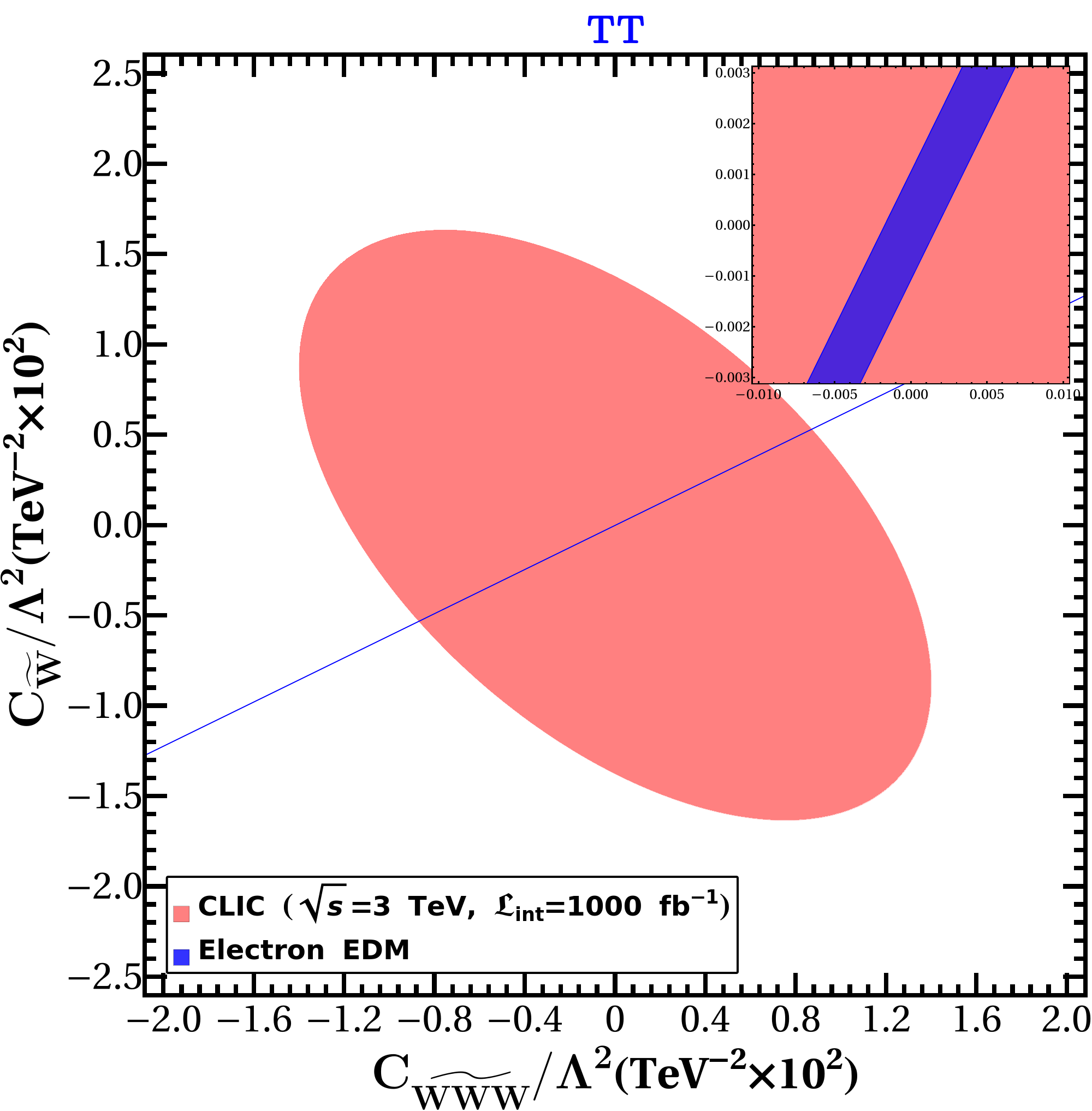}~	
	\includegraphics[height=5cm,width=4.95cm]{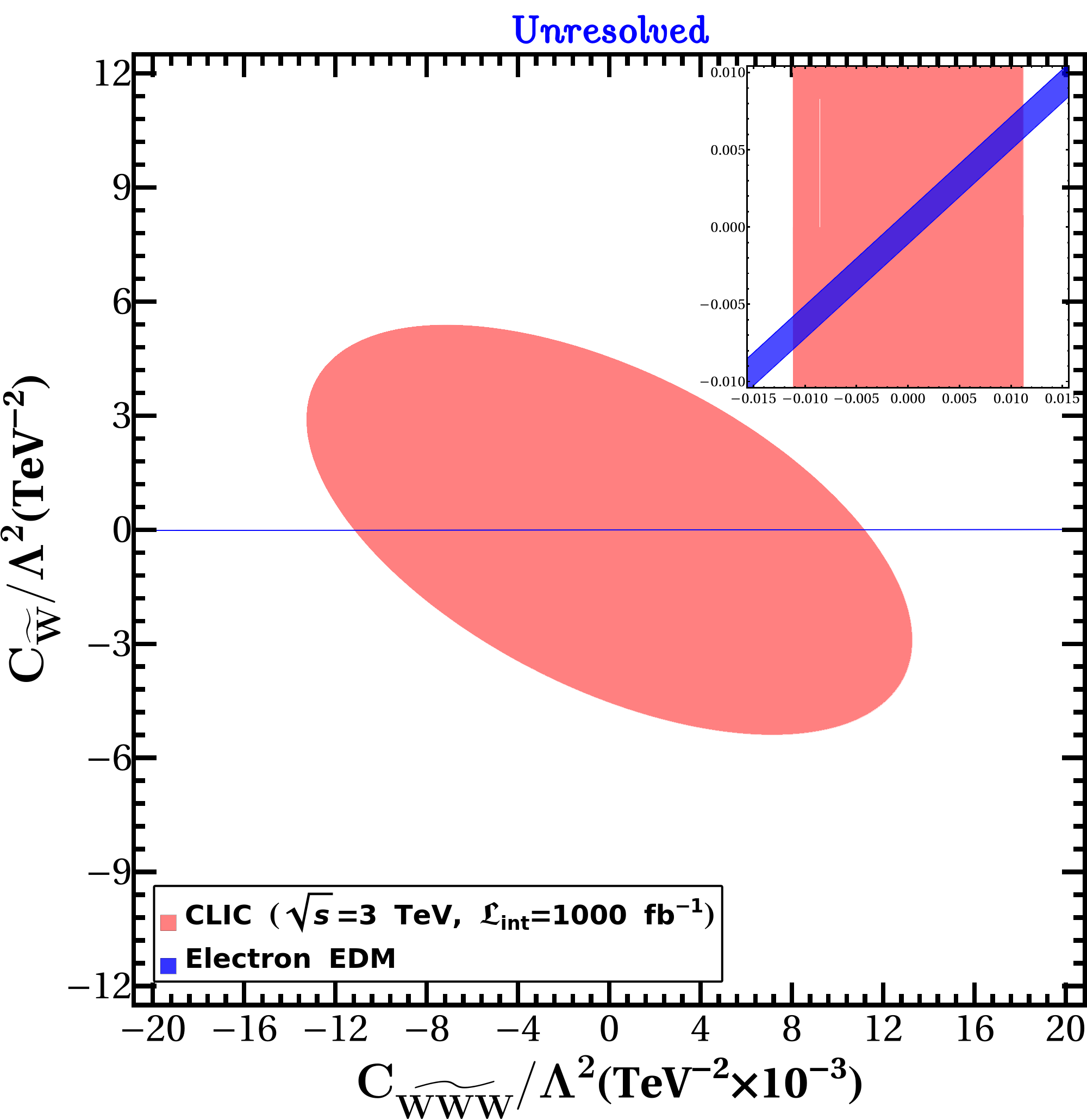}
	$$
	\caption{Electron EDM (Blue color) and CLIC sensitivity (Pink color) in $C_{\widetilde{WWW}}/\Lambda^2-C_{\widetilde{W}}/\Lambda^2$ plane. For CLIC, we consider $\{P_{e^-}:P_{e^+}=+80\%:0\%\}$ polarization combination. The blue regions in all the plots are allowed by the electron EDM. Please note the different scaling along x and y axes in all the figures.} 
	\label{fig:cpv.2d}
\end{figure}

\noindent
where $y_e$, $s_w$, $m_w$, $m_h$, and $\Lambda_C$ are the electron-Yukawa coupling, sine of the weak mixing angle, $W$-boson mass, Higgs boson mass, and cut-off scale, respectively. Using Eqs.~\eqref{eq:edm.cons} and \eqref{eq:edm},  we obtain the following limits (considering one operator at a time) on CP-violating couplings as 
\begin{align}
	\begin{split}
		|C_{\widetilde{WWW}}/\Lambda^2|  & \lesssim 0.002~\text{TeV}^{-2} ,\\
		|C_{\widetilde{W}}/\Lambda^2|  & \lesssim 0.001~\text{TeV}^{-2}.
	\end{split}
	\label{eq:edm.limit}
\end{align}
\noindent
Comparing the bounds on CP-violating operators from the electron EDM (Eq.~\eqref{eq:edm.limit}) with the most stringent sensitivity in Table~\ref{tab:oot.limit} obtainable at the CLIC, we find that the limits are similar for $C_{\widetilde{WWW}}/\Lambda^2$. However, in case of $C_{\widetilde{W}}/\Lambda^2$, the EDM constraint is three orders of magnitude stronger than the corresponding limit achievable at CLIC with $\sqrt{s}=3$ TeV and $\mathfrak{L}_{\text{int}}=1000~\text{fb}^{-1}$. In Fig.~\ref{fig:cpv.2d}, we present the electron EDM bound and CLIC sensitivity in $C_{\widetilde{WWW}}/\Lambda^2-C_{\widetilde{W}}/\Lambda^2$ plane, considering the contribution from both operators simultaneously. The dominance of the operator $\mathscr{O}_{\widetilde{WWW}}$ in the collider scenario is evident from the figure for `unresolved' and `TL+LT' helicity combination of $W$ boson pair, whereas, for TT helicity combination, both operators contribute to the similar order. For the electron EDM scenario, the contributions from both operators are of the same order of magnitude.
%
\section{Conclusion}
\label{sec:con}
\noindent
Precise measurement of charged triple gauge couplings (cTGCs) is essential to test the SM and look for potential signals beyond the Standard Model (BSM). In this paper, we have analyzed the precision measurement of cTGCs through $WW$ production process followed by semi-leptonic decay in the upcoming electron-positron collider with the initial polarized beams. Due to the significant cross~section, the $WW$ process benefits from a sufficient event rate to perform precision measurement of cTGCs at $\sqrt{s}=3$ TeV and $\mathfrak{L}_{\text{int}}=1000$ fb$^{-1}$. We have parameterized the deviation of cTGCs in the SMEFT framework with three CP-even and two CP-odd dimension-6 SMEFT operators. We then performed cut-based analysis to estimate signal-background separation using different kinematical variables. A machine learning technique, Boosted Decision Tree (BDT) analysis, has also been performed to improve signal and background separation.

We have employed the optimal observable technique (OOT) to estimate the sensitivity of dimension-6 cTGCs considering differential distribution as an observable with different helicity combinations of $W$ boson pair. We have established that optimal sensitivities of NP couplings are improved by two orders of magnitude for the CP-conserving case and one order of magnitude for the CP-violating case with respect to the existing LHC bounds on the respective couplings. Furthermore, we have compared the projected sensitivity at CLIC using standard $\chi^2$-analysis and OOT. We have found that compared to the standard $\chi^2$-square analysis, the optimal sensitivities are significantly stringent, highlighting the effectiveness of the OOT. Different helicity combinations of $W$ boson pairs exhibit varying sensitivity for different dimension-6 cTGCs, offering a pathway to probe specific NP couplings. Initial beam polarizations play a critical role in enhancing the sensitivity of NP couplings as well, influencing their estimation by several factors. Two CP-odd effective operators contribute to the electric dipole moment (EDM) at the one-loop level. Our analysis reveals that, given the center-of-mass (CM) energy and luminosity parameters mentioned earlier, the sensitivity obtained at the future electron-positron collider experiments for the operator $\mathscr{O}_{\widetilde{WWW}}$ is comparable to that of the electron EDM. However, for $\mathscr{O}_{\widetilde{W}}$, the current EDM experiment imposes more stringent constraints than those achievable at colliders. 

\acknowledgments
This work is partially supported by the Guangdong Major Project of Basic and Applied Basic Research, Grant No. 2020B0301030008. SJ thanks Xiao-Dong Ma for useful discussions.

\appendix

\section{Wigner's functions}
\label{sec:wig.func}
Wigner's functions in Eq.~\eqref{eq:amp.ww} are expressed as \cite{Hagiwara:1986vm}
\begin{align}
	\begin{split}
		d^{2}_{1,2}=-d^{2}_{-1,-2}&=\frac{1}{2}(1+\cos\theta)\sin \theta,\\
		d^{2}_{1,-2}=-d^{2}_{-1,2}&=-\frac{1}{2}(1-\cos\theta)\sin \theta,\\
		d^{1}_{1,1}=-d^{1}_{-1,-1}&=\frac{1}{2}(1+\cos\theta),\\
		d^{1}_{1,-1}=-d^{1}_{-1,1}&=\frac{1}{2}(1-\cos\theta),\\
		d^{1}_{1,0}=-d^{1}_{-1,0}&=\frac{1}{\sqrt{2}}\sin \theta.\\
	\end{split}
	\label{eq:wig.func}
\end{align}

\section{Optimal sensitivity comparison: CEPC vs FCC-ee vs CLIC}
\label{sec:sen_comp}
The sensitivities of the NP couplings at the Circular Electron-Positron Collider (CEPC) with $\sqrt{s} = 240$~GeV~\cite{CEPCStudyGroup:2023quu} and the Future Circular Electron-Positron Collider (FCC-ee) with $\sqrt{s} = 365$~GeV~\cite{Agapov:2022bhm}, both assuming an integrated luminosity of $\mathfrak{L}_{\text{int}} = 1000~\text{fb}^{-1}$ and unpolarized beam, are presented in Table~\ref{tab:col.sen.comp}. We consider the unresolved helicity of the $WW$ state for the sensitivity comparison. These sensitivities are significantly weaker compared to those at the CLIC, primarily because at lower CM energies, the SM contribution to $WW$ production is larger, while the SMEFT contribution is relatively suppressed.
\begin{table}[H]
	\centering
	{\renewcommand{\arraystretch}{1.3}%
		\begin{tabular}{ |c|c|c|c|c| } 
			\hline
			NP couplings & CEPC & FCC-ee & CLIC \\ 
			\hline
			$C_{B}/\Lambda^2$ & [$-344.80,+139.20$] & [$-107.06,+39.41$]  & $[-0.86,+0.11]$\\
			$C_{W}/\Lambda^2$ & [$-144.72,+50.30$] & $[-57.71,+18.23]$ & $[-0.63,+0.06]$\\ 
			$C_{WWW}/\Lambda^2$ & [$-97.57,+55.18$] & $[-26.76,+18.76]$ & $[-0.11,+0.10]$\\ 
			$C_{\widetilde{WWW}}/\Lambda^2$ & [$-55.78,+55.78$] & $[-12.14,+12.14]$ & $[-0.006,+0.006]$\\ 
			$C_{\widetilde{W}}/\Lambda^2$ & [$-240.32,+240.32$] & $[-120.34,+120.34]$ & $[-3.84,+3.84]$\\
			\hline
	\end{tabular}}
	\caption{Comparison of optimal sensitivity (95\% C.L.) among CEPC ($\sqrt{s}=240$ GeV), FCC-ee ($\sqrt{s}=365$ GeV), and CLIC ($\sqrt{s}=3$ TeV) with $\mathfrak{L}_{\text{int}}=1000~\text{fb}^{-1}$ and unpolarized beam for unresolved $WW$ helicity state.}
	\label{tab:col.sen.comp}
\end{table}
\bibliographystyle{JHEP}
\bibliography{ref.bib}
\end{document}

%% file: mylatex.tex
\usepackage{bbm}  
\usepackage{amsmath}                                                

\newcommand{\nc}{\newcommand}
\nc{\beq}{\begin{equation}}  \nc{\eeq}{\end{equation}}
\nc{\bea}{\begin{eqnarray}}  \nc{\eea}{\end{eqnarray}}
\nc{\baa}{\begin{array}}     \nc{\eaa}{\end{array}}
\nc{\bit}{\begin{itemize}}   \nc{\eit}{\end{itemize}}
\nc{\ben}{\begin{enumerate}} \nc{\een}{\end{enumerate}}
\nc{\bce}{\begin{center}}    \nc{\ece}{\end{center}}
\nc{\bpm}{\begin{pmatrix}}   \nc{\epm}{\end{pmatrix}}
\nc{\bvt}{\begin{verbatim}}  \nc{\evt}{\end{verbatim}}
\nc{\bal}{\begin{align}}
\def\to{\rightarrow}

\def\boldoverdot{\,{\raise6pt\hbox{\bf.}\!\!\!\!\>}}

\def\ee{{\bf e}}

\usepackage{bm}
\def\diag{\hbox{\diag}}

\def\doubleundertext#1{
{\undertext{\vphantom{y}#1}}\par\nobreak\vskip-\the\baselineskip\vskip4pt%
\undertext{\hbox to 2in{}}}
\def\inbox#1{\vbox{\hrule\hbox{\vrule\kern5pt
     \vbox{\kern5pt#1\kern5pt}\kern5pt\vrule}\hrule}}
\def\sqr#1#2{{\vcenter{\hrule height.#2pt
      \hbox{\vrule width.#2pt height#1pt \kern#1pt
         \vrule width.#2pt}
      \hrule height.#2pt}}}

\def\today{\ifcase\month\or
  January\or February\or March\or April\or May\or June\or
  July\or August\or September\or October\or November\or December\fi
  \space\number\day, \number\year}
\def\pmb#1{\setbox0=\hbox{#1}%
  \kern-.025em\copy0\kern-\wd0
  \kern.05em\copy0\kern-\wd0
  \kern-.025em\raise.0433em\box0 }

\def\pmbb#1{\setbox0=\hbox{#1}%
  \kern-.02em\copy0\kern-\wd0
  \kern.04em\copy0\kern-\wd0
  \kern-.02em\raise.03464em\box0 }
\def\sumprime_#1{\setbox0=\hbox{$\scriptstyle{#1}$}
  \setbox2=\hbox{$\displaystyle{\sum}$}
  \setbox4=\hbox{${}'\mathsurround=0pt$}
  \dimen0=.5\wd0 \advance\dimen0 by-.5\wd2
  \ifdim\dimen0>0pt
  \ifdim\dimen0>\wd4 \kern\wd4 \else\kern\dimen0\fi\fi
\mathop{{\sum}'}_{\kern-\wd4 #1}}